\documentclass[prd,aps,showpacs,nofootinbib,preprint,eqsecnum]{revtex4}
\usepackage{graphicx,color,amsmath,amsxtra}
\usepackage{epsf}
\usepackage{amssymb}
\usepackage{enumerate}
\usepackage{hhline}
\usepackage{array}
\usepackage{tabularx}

\newcommand{\bear}{\begin{array}}  \newcommand{\eear}{\end{array}}
\newcommand{\bea}{\begin{eqnarray}}  \newcommand{\eea}{\end{eqnarray}}
\newcommand{\beq}{\begin{equation}}  \newcommand{\eeq}{\end{equation}}
\newcommand{\bef}{\begin{figure}}  \newcommand{\eef}{\end{figure}}
\newcommand{\bec}{\begin{center}}  \newcommand{\eec}{\end{center}}

  \newcommand{\abs}[1]{\vert{#1}\vert}

\newcommand{\Eqn}[1]{&\hspace{-0.2em}#1\hspace{-0.2em}&}

\def\Vec#1{\mbox{\boldmath $#1$}}

\def\be{\begin{equation}}
\def\ee{\end{equation}}
\def\bea{\begin{eqnarray}}
\def\eea{\end{eqnarray}}
\def\beq{\begin{eqnarray}}
\def\eeq{\end{eqnarray}}

\def\be{\begin{equation}}
\def\ee{\end{equation}}
\def\bea{\begin{eqnarray}}
\def\eea{\end{eqnarray}}
\def\beq{\begin{eqnarray}}
\def\eeq{\end{eqnarray}}

\begin{document}

\title{Equation of state for dark energy in $f(T)$ gravity}

\author{Kazuharu Bamba\footnote{
E-mail address: 
bamba@kmi.nagoya-u.ac.jp;~
\vspace{-0.4em}
Present address: 
Kobayashi-Maskawa Institute for the Origin of Particles and the Universe, 
Nagoya University, Nagoya 464-8602, Japan.}, 
Chao-Qiang Geng\footnote{E-mail address: geng@phys.nthu.edu.tw}, 
Chung-Chi Lee\footnote{E-mail address: g9522545@oz.nthu.edu.tw} 
and 
Ling-Wei Luo\footnote{E-mail address: d9622508@oz.nthu.edu.tw} 
}
\affiliation{
Department of Physics, National Tsing Hua University, Hsinchu, Taiwan 300 
}


\begin{abstract}

We study the cosmological evolutions of the equation of state for dark energy 
$w_{\mathrm{DE}}$ in the exponential and logarithmic 
as well as their combination $f(T)$ theories. 
We show that the crossing of the phantom divide line of 
$w_{\mathrm{DE}} = -1$ can be realized in the combined $f(T)$ theory 
even though it cannot be in the exponential 
or logarithmic $f(T)$ theory. 
In particular, the crossing is from $w_{\mathrm{DE}} > -1$ to 
$w_{\mathrm{DE}} < -1$, in the opposite manner from $f(R)$ gravity models. 
We also demonstrate that this feature is favored 
by the recent observational data. 

\end{abstract}

\pacs{
98.80.-k, 04.50.Kd
}

\maketitle

\section{Introduction}

Cosmic observations from Supernovae Ia (SNe Ia)~\cite{SN1}, 
cosmic microwave background (CMB) radiation~\cite{WMAP, Komatsu:2008hk, 
Komatsu:2010fb}, 
large scale structure (LSS)~\cite{LSS},  
baryon acoustic oscillations (BAO)~\cite{BAO}, 
and weak lensing~\cite{WL} 
have implied that the expansion of the universe is currently accelerating. 
This is one of the most important issues in modern physics. 
Approaches to account for the late time cosmic acceleration 
fall into two representative categories: 
One is to introduce ``dark energy'' in 
the right-hand side of the Einstein equation 
in the framework of general relativity 
(for a review on dark energy, see~\cite{Copeland:2006wr}). 
The other is to modify the left-hand side of the Einstein equation, called as 
a modified gravitational theory, 
e.g., $f(R)$ gravity~\cite{Review-N-O, Sotiriou:2008rp, DeFelice:2010aj}. 

As another possible way to examine gravity beyond general relativity, 
one could use the Weitzenb\"{o}ck connection, which has no curvature but 
torsion, rather than the curvature defined by the Levi-Civita connection. 
Such an approach is referred to ``teleparallelism" (see, 
e.g.,~\cite{Hehl:1976kj, Hayashi:1979qx, Flanagan:2007dc, Garecki:2010jj}), 
which was also taken by Einstein~\cite{Einstein}. 
To explain the late time acceleration of the universe, 
the teleparallel Lagrangian density described by the torsion scalar $T$ 
has been extended to a function of $T$~\cite{Bengochea:2008gz, 
Linder:2010py}\footnote{Models based on modified teleparallel gravity 
for inflation have been investigated in Ref.~\cite{F-F}.}. 
This idea is equivalent to the concept of $f(R)$ gravity, in which 
the Ricci scalar $R$ in the Einstein-Hilbert action is promoted to 
a function of $R$. 
Recently, $f(T)$ gravity has been extensively studied in the 
literature~\cite{f(T)-Refs, Chen:2010va, Wu:2010av, BGL-Comment, 
Local-Lorentz-invariance, Dent:2010bp}. 

In this paper, we explicitly examine the cosmological evolution in 
the exponential $f(T)$ theory~\cite{Linder:2010py, BGL-Comment} 
in more detail with the analysis method in Ref.~\cite{Hu:2007nk}. 
In particular, we study the equation of state ($w_{\mathrm{DE}}$) 
and energy density ($\rho_{\mathrm{DE}}$) for dark energy. 
The recent cosmological observational data~\cite{observational status} seems 
to imply a dynamical dark energy of equation of state with 
the crossing of the phantom divide line $w_{\mathrm{DE}}=-1$ 
from the non-phantom phase to phantom phase as the redshift $z$ decreases
in the near past. 
However, 
we illustrate that 
the universe with the exponential $f(T)$ theory always stays in the non-phantom (quintessence) phase 
or the phantom one, 
and hence the crossing of the phantom divide 
cannot be realized~\cite{BGL-Comment}. 
It is interesting to mention that 
such an exponential type as $f(R)$ gravity models
has been investigated in Refs.~\cite{Exponential-type-f(R)-gravity, B-G-L, 
Yang:2010xq}. 
We also present a logarithmic $f(T)$ theory 
and show that it has a similar feature as the exponential one. 
Our motivation in this paper is to build up a realistic $f(T)$ theory 
in which the same behavior of
the crossing of the phantom divide as indicated by the data can be achieved. 
For this purpose,
we will construct a combined $f(T)$ theory with both 
logarithmic and exponential terms. 
Furthermore, we examine the observational constraints on the 
combined $f(T)$ theory by using the recent 
observational data of SNe Ia, BAO and CMB. 
We note that 
two $f(T)$ models with the crossing of the phantom divide 
have also been proposed in Ref.~\cite{Wu:2010av}. 

The paper is organized as follows. 
In Sec.\ II, we introduce $f(T)$ theory and derive the gravitational field 
equations. 
In Sec.\ III, we investigate the cosmological evolutions in the exponential 
and logarithmic $f(T)$ theories. 
In Sec.\ IV, 
we propose a model which combines the two theories in Sec.\ III 
to achieve the crossing of the phantom divide. 
In Sec.\ V, we explore the observational constraints on the combined model 
in Sec.\ IV. 
Finally, conclusions are given in Sec.\ VI. 

\section{The cosmological formulae in $f(T)$ gravity}

In the teleparallelism, 
orthonormal tetrad components $e_A (x^{\mu})$ are used, 
where an index $A$ runs over $0, 1, 2, 3$ for the 
tangent space at each point $x^{\mu}$ of the manifold. 
Their relation to the metric $g^{\mu\nu}$ is given by 
\begin{equation}
g_{\mu\nu}=\eta_{A B} e^A_\mu e^B_\nu\,, 
\label{eq:2.1}
\end{equation}
where $\mu$ and $\nu$ are coordinate indices on the manifold 
and also run over $0, 1, 2, 3$, 
and $e_A^\mu$ forms the tangent vector of the manifold. 

The torsion $T^\rho_{\verb| |\mu\nu}$ and contorsion 
$K^{\mu\nu}_{\verb|  |\rho}$ tensors are defined by
\begin{eqnarray}
T^\rho_{\verb| |\mu\nu} \Eqn{\equiv} e^\rho_A 
\left( \partial_\mu e^A_\nu - \partial_\nu e^A_\mu \right)\,, 
\label{eq:2.2} \\ 
K^{\mu\nu}_{\verb|  |\rho} 
\Eqn{\equiv} 
-\frac{1}{2} 
\left(T^{\mu\nu}_{\verb|  |\rho} - T^{\nu \mu}_{\verb|  |\rho} - 
T_\rho^{\verb| |\mu \nu}\right)\,.
\label{eq:2.3}
\end{eqnarray}
Instead of the Ricci scalar $R$ for the Lagrangian density 
in general relativity, 
the teleparallel Lagrangian density is described by the torsion scalar 
$T$, defined as 
\begin{equation}
T \equiv S_\rho^{\verb| | \mu\nu} T^\rho_{\verb| | \mu\nu}\,,
\label{eq:2.4}
\end{equation}
where
\begin{equation}
S_\rho^{\verb| | \mu\nu} \equiv \frac{1}{2}
\left(K^{\mu\nu}_{\verb|  |\rho}+\delta^\mu_\rho \ T^{\alpha \nu}_{\verb|  | \alpha} - \delta ^\nu _ \rho \ 
T ^{\alpha \mu }_{\verb|  | \alpha }\right)\,. 
\label{eq:2.5}
\end{equation}
Consequently, the modified teleparallel action for $f(T)$ theory 
is given by~\cite{Linder:2010py} 
\begin{equation}
I=\frac{1}{16 \pi G}\int d^4x \abs{e} \left( T + f(T) \right)\,,
\label{eq:2.6}
\end{equation}
where $\abs{e}= \det \left(e^A_\mu \right)=\sqrt{-g}$. 

We assume the flat Friedmann-Lema\^{i}tre-Robertson-Walker (FLRW) 
space-time with the metric, 
\begin{equation}
{ds}^2 = {dt}^2 - a^2(t)d{\Vec{x}}^2\,,
\label{eq:2.7}
\end{equation}
where $a(t)$ is the scale factor. 
In this space-time, 
$g_{\mu \nu}= \mathrm{diag} (1, -a^2, -a^2, -a^2)$ and therefore 
the tetrad components $e^A_\mu = (1,a,a,a)$ 
yield the exact value of torsion scalar $T=-6H^2$, 
where $H=\dot{a}/a$ is the Hubble parameter. 
We use units of $k_\mathrm{B} = c = \hbar = 1$ and 
the gravitational constant 
$G = M_{\mathrm{Pl}}^{-2}$ 
with the Planck mass of 
$M_{\mathrm{Pl}} = 1.2 \times 10^{19}$\,\,GeV. 

It follows from the variation principle that 
in the flat FLRW background, 
the modified Friedmann equations are given 
by~\cite{Bengochea:2008gz, Linder:2010py} 
\begin{eqnarray}
H^2 \Eqn{=} \frac{8 \pi G }{3}\rho_{\mathrm{M}} -\frac{f}{6}-2H^2 f_T \,,
\label{eq:2.8} \\ 
\left(H^2\right)^\prime \Eqn{=} 
\frac{16 \pi GP_{\mathrm{M}} 
+ 6H^2+f+12H^2f_T}{24H^2f_{TT}-2-2f_T}\,, 
\label{eq:2.9}
\end{eqnarray}
where a prime denotes a derivative with respect to $\ln a$, 
$f_T \equiv df(T)/dT$, $f_{TT} \equiv d^2f(T)/dT^2$, 
and $\rho_{\mathrm{M}}$ and $P_{\mathrm{M}}$ are 
the energy density and pressure of all perfect fluids of generic matter, 
respectively. 

By comparing the above modified Friedmann equations (\ref{eq:2.8}) and 
(\ref{eq:2.9}) with the ordinary ones in general relativity: 
\begin{eqnarray}
H^2 \Eqn{=} \frac{8 \pi G }{3} \left(\rho_{\mathrm{M}}+\rho_{\mathrm{DE}} 
\right)\,, 
\label{eq:2.10} \\ 
\left(H^2\right)^\prime 
\Eqn{=} -8 \pi G \left(\rho_{\mathrm{M}} + P_{\mathrm{M}} + 
\rho_{\mathrm{DE}} + P_{\mathrm{DE}} \right)\,,
\label{eq:2.11} 
\end{eqnarray}
the energy density and pressure of the effective dark energy can be 
described by
\begin{eqnarray}
\rho_{\mathrm{DE}} \Eqn{=} \frac{1}{16 \pi G}\left(-f+2Tf_{T}\right)\,,
\label{eq:2.12} \\ 
P_{\mathrm{DE}} \Eqn{=} 
\frac{1}{16 \pi G} \frac{f-Tf_T+2T^2f_{TT}}{1+f_T+2Tf_{TT}}\,.
\label{eq:2.13} 
\end{eqnarray}
The equation of state for dark energy is defined as~\cite{Linder:2010py} 
\begin{equation}
w_{\mathrm{DE}} \equiv \frac{P_{\mathrm{DE}}}{\rho_{\mathrm{DE}}} 
= 
-1+\frac{T^\prime}{3T}\frac{f_T+2Tf_{TT}}{f/T-2f_T}=-\frac{f/T-f_T+2Tf_{TT}}{
\left(1+f_T+2Tf_{TT}\right)\left(f/T-2f_T\right)}\,.
\label{eq:2.14} 
\end{equation}
%
Since we are interested in the late time universe, 
we consider only non-relativistic matter (cold dark matter and baryon) 
with $\rho_{\mathrm{M}} = \rho_{\mathrm{m}}$ and 
$P_{\mathrm{M}} = P_{\mathrm{m}} = 0$, 
where $\rho_{\mathrm{m}}$ and $P_{\mathrm{m}}$ are the energy density and 
pressure of non-relativistic matter, respectively. 
Consequently, from Eqs.~(\ref{eq:2.8}) and (\ref{eq:2.9}) 
it can be shown that the effective dark energy satisfies
the continuity equation 
\begin{equation}
\frac{d \rho_{\mathrm{DE}}}{dN} 
\equiv \rho_{\mathrm{DE}}^\prime = -3 
\left(1+w_{\mathrm{DE}}\right) \rho_{\mathrm{DE}}\,, 
\label{eq:2.15} 
\end{equation}
where $N \equiv \ln a$.

\section{Cosmological evolution in the exponential 
and logarithmic $f(T)$ theories}

To analyze the cosmological evolution of 
the equation of state for dark energy 
$w_{\mathrm{DE}}$ in $f(T)$ theory, 
we define a dimensionless variable~\cite{Hu:2007nk}: 
\begin{equation}
y_H \equiv 
\frac{H^2}{\bar{m}^2}-a^{-3}= 
\frac{\rho_{\mathrm{DE}}}{\rho_{\mathrm{m}}^{(0)}}\,,
\label{eq:3.1} 
\end{equation}
where
\begin{equation}
\bar{m}^2 \equiv \frac{8 \pi G \rho_{\mathrm{m}}^{(0)}}{3}\,,
\label{eq:3.2} 
\end{equation}
and $\rho_{\mathrm{m}}^{(0)} = \rho_{\mathrm{m}}(z=0)$ is the current density 
parameter of non-relativistic matter with the redshift $z \equiv 1/a -1$. 
Using Eq.~(\ref{eq:2.15}), we obtain the evolution equation of the universe as 
follows: 
\begin{equation}
y_H^\prime = -3\left(1+w_{\mathrm{DE}}\right)y_H\,.
\label{eq:3.3} 
\end{equation}
Note that 
$w_{\mathrm{DE}}$ is a function of $T$, while
$T$ is a function of $H^2$. 
Moreover, 
it follows from Eq.~(\ref{eq:3.1}) that 
$H^2=\bar{m}^2 \left(y_H + a^{-3}\right)$.

\subsection{Exponential $f(T)$ theory}

We consider the exponential $f(T)$ theory in Ref.~\cite{Linder:2010py}, 
given by 
\begin{equation}
f(T)= \alpha T \left(1-e^{pT_0/T}\right)
\label{eq:3.4} 
\end{equation}
with
\begin{equation}
\alpha = -\frac{1-\Omega_{\mathrm{m}}^{(0)}}{1-\left(1-2p\right)e^p}\,, 
\label{eq:3.5} 
\end{equation}
where $p$ is a constant with $p=0$ corresponding to the $\Lambda$CDM model
and
$T_0=T(z=0)$ is the current torsion.
Here, 
$\Omega_{\mathrm{m}}^{(0)} \equiv 
\rho_{\mathrm{m}}^{(0)}/\rho_{\mathrm{crit}}^{(0)}$ 
where 
$\rho_{\mathrm{m}}^{(0)}$
is the energy density of non-relativistic matter 
at the present time and 
$\rho_{\mathrm{crit}}^{(0)} = 3H_0^2/\left( 8 \pi G \right)$ is the critical 
density with $H_0$ being the current Hubble parameter. 
We note that 
$\alpha$ in Eq.~(\ref{eq:3.5}) has been derived from
$
\Omega_{\mathrm{DE}}^{(0)} 
\equiv \rho_{\mathrm{DE}}^{(0)}/\rho_{\mathrm{crit}}^{(0)}
= 1-\Omega_{\mathrm{m}}^{(0)} 
= -\alpha \left[ 1-\left(1-2p\right)e^p \right]
$ 
by using
Eq.~(\ref{eq:2.12}) with 
$T=T_0$ and
$\rho_{\mathrm{DE}}^{(0)} = \rho_{\mathrm{DE}} (z=0)$.
We remark that the theory in Eq.~(\ref{eq:3.4}) contains only one parameter 
$p$ if the value of $\Omega_{\mathrm{m}}^{(0)}$ is given. 
The values of other dimensionless quantities at $z=0$ can be calculated 
by using $p$ and $\Omega_{\mathrm{m}}^{(0)}$, e.g., 
$\bar{m}^2/T_0
=\left(8 \pi G \rho_{\mathrm{m}}^{(0)}/3\right)/
\left(-6H_0^2\right) 
=-\Omega_{\mathrm{m}}^{(0)}/6$ 
and 
$y_{H}(z=0)=H_0^2/\bar{m}^2-1=1/\Omega_{\mathrm{m}}^{(0)}-1$.

\begin{center}
\begin{figure}[tbp]
\begin{tabular}{ll}
\begin{minipage}{80mm}
\begin{center}
\unitlength=1mm
\resizebox{!}{6.5cm}{
   \includegraphics{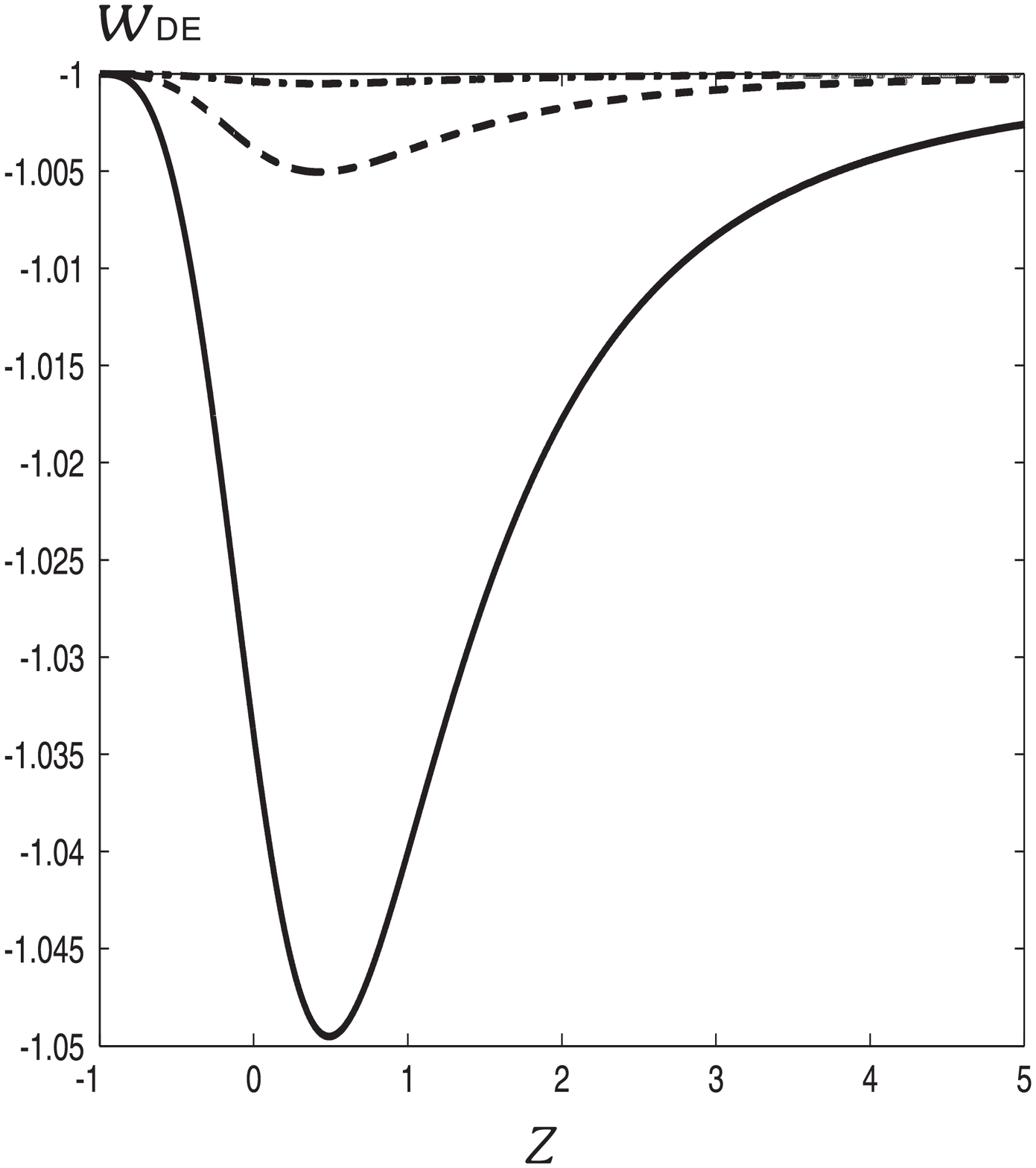}
                  }
\end{center}
\end{minipage}
&
\begin{minipage}{80mm}
\begin{center}
\unitlength=1mm
\resizebox{!}{6.5cm}{
   \includegraphics{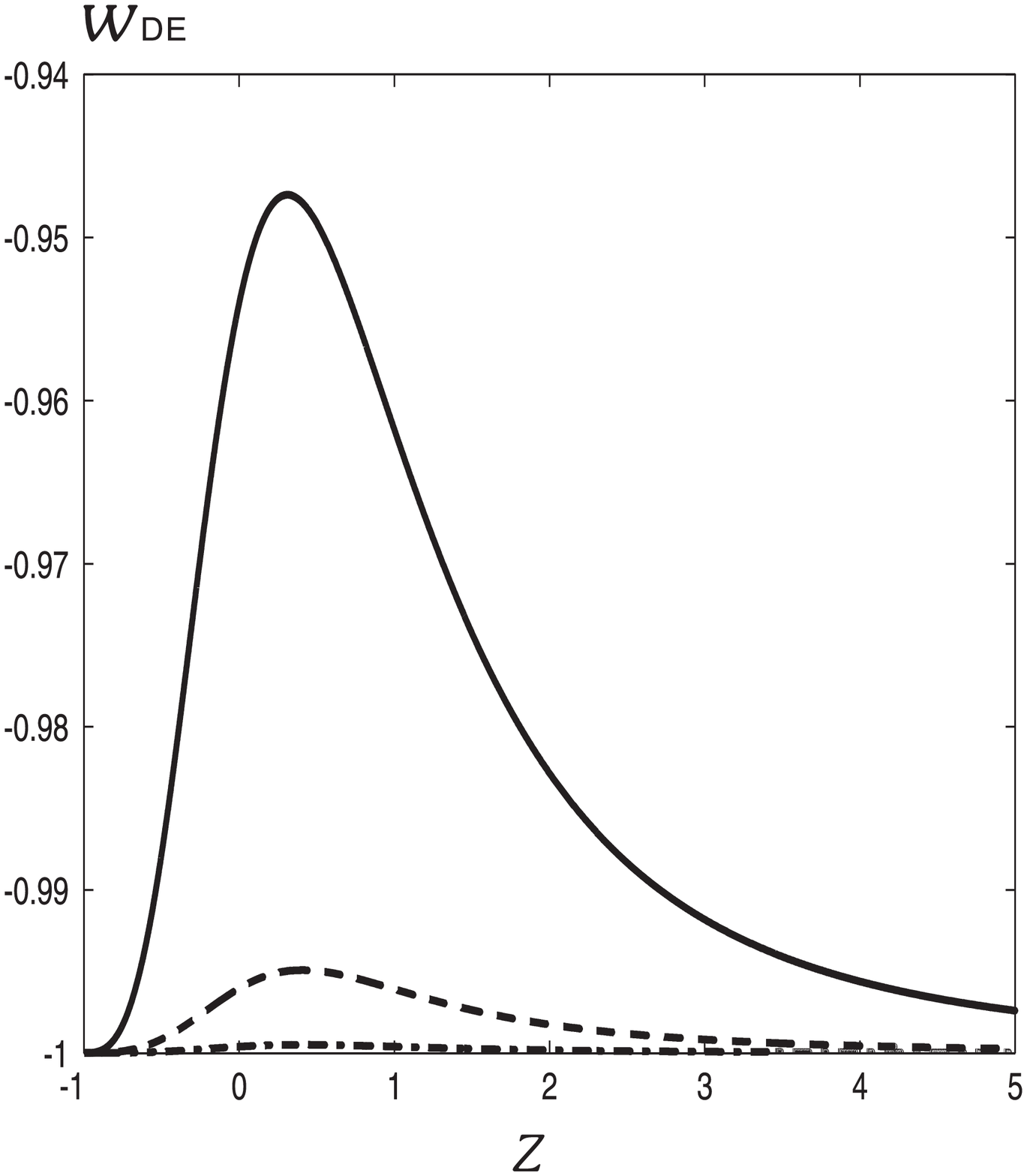}
                  }
\end{center}
\end{minipage}

\end{tabular}
\caption{$w_{\mathrm{DE}}$ 
as a function of the redshift $z$ for $\abs{p}=0.1$ (solid line), 
$0.01$ (dashed line), $0.001$ (dash-dotted line) and 
$\Omega_{\mathrm{m}}^{(0)} = 0.26$ 
in the exponential $f(T)$ theory, where 
the left and right panels are for $p>0$ and $p<0$, respectively. 
}
\label{fig-1}
\end{figure}
\end{center}


In Fig.~\ref{fig-1}, we depict the equation of state for dark energy 
$w_{\mathrm{DE}}$ in Eq.~(\ref{eq:2.14}) 
as a function of the redshift $z$ for $\abs{p}=0.1$, $0.01$, $0.001$. 
We have used 
$\Omega_{\mathrm{m}}^{(0)} = 0.26$~\cite{Komatsu:2010fb} 
in all numerical calculations. 
{}From Fig.~\ref{fig-1}, we see that $w_{\mathrm{DE}}$ does not 
cross the phantom divide line $w_{\mathrm{DE}}=-1$ in 
the exponential $f(T)$ theory~\cite{Linder:2010py, BGL-Comment}. 
In particular, 
for $p<0$ the universe always stays in the non-phantom (quintessence) phase 
($w_{\mathrm{DE}} > -1$), 
whereas for $p>0$ it in the phantom phase 
($w_{\mathrm{DE}} < -1$). 
The present values of $w_{\mathrm{DE}}$ are 
$w_{\mathrm{DE}} (z=0) = -1.03$, $-1.003$, $-1.0003$, 
$-0.954$, $-0.996$ and $-0.999$ 
for 
$p=0.1$, $0.01$, $0.001$, 
$p=-0.1$, $-0.01$ and $-0.001$, 
respectively. 
The larger $\abs{p}$ is, the larger 
the deviation of the exponential $f(T)$ theory 
from the $\Lambda\mathrm{CDM}$ model is. 
We note that in solving Eq.~(\ref{eq:3.3}) numerically, 
we have taken the initial conditions 
at $z=0$ as $y_{H}(z=0) = 2.8$. 


\begin{center}
\begin{figure}[tbp]
\begin{tabular}{ll}
\begin{minipage}{80mm}
\begin{center}
\unitlength=1mm
\resizebox{!}{6.5cm}{
   \includegraphics{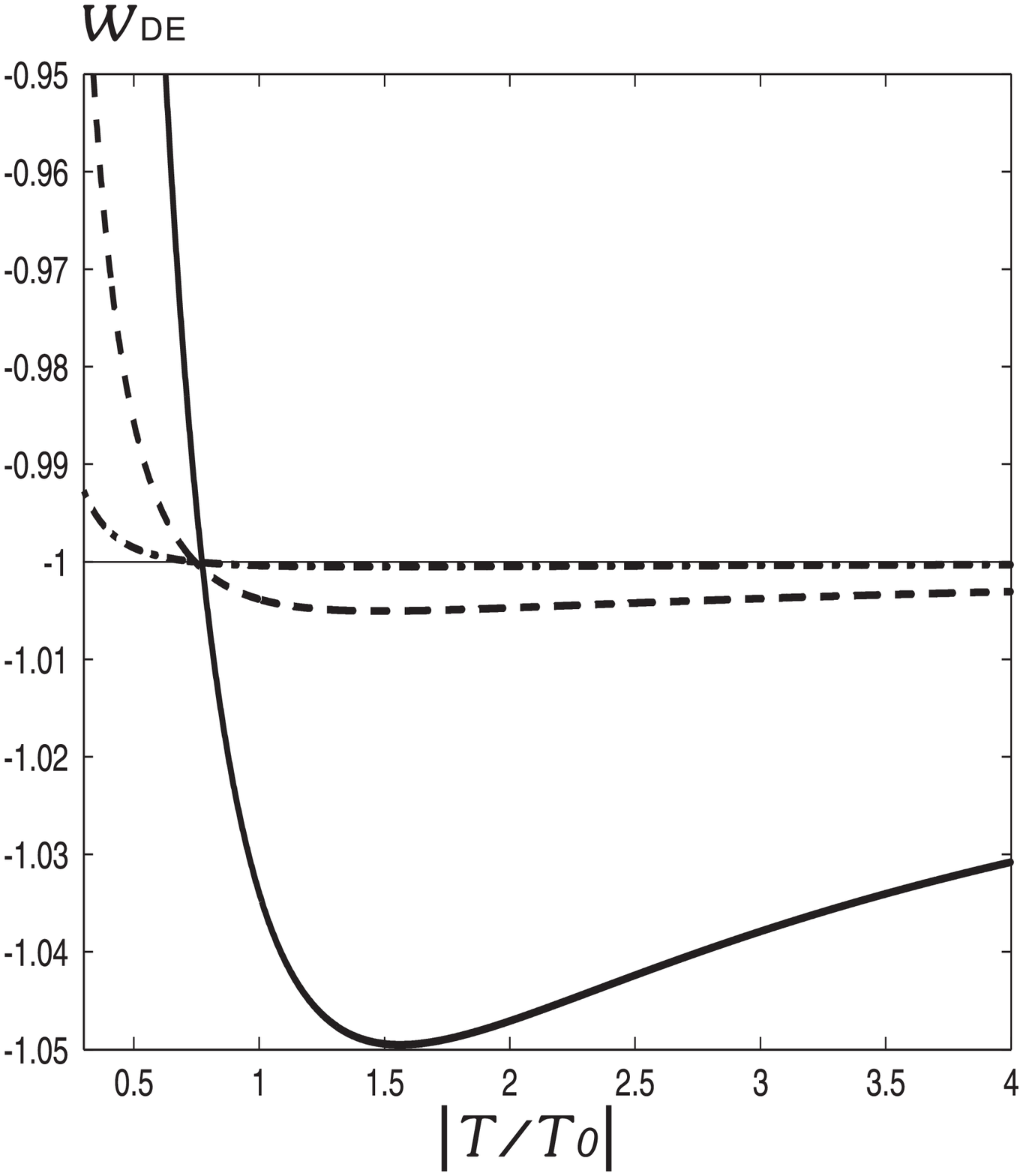}
                  }
\end{center}
\end{minipage}
&
\begin{minipage}{80mm}
\begin{center}
\unitlength=1mm
\resizebox{!}{6.5cm}{
   \includegraphics{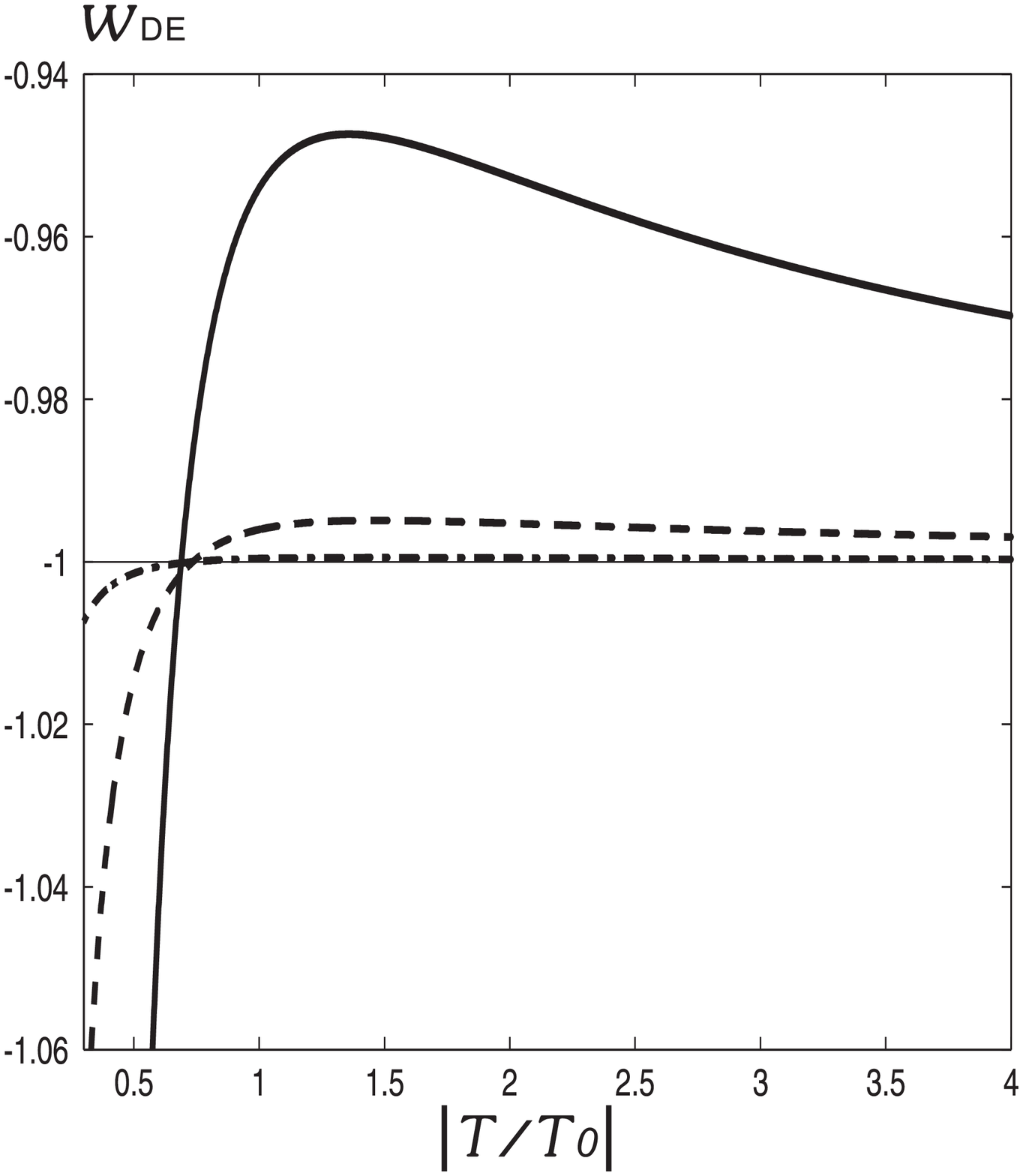}
                  }
\end{center}
\end{minipage}

\end{tabular}
\caption{$w_{\mathrm{DE}}$ 
as a function of $\abs{ T/T_0 }$, 
where the thin solid line shows $w_{\mathrm{DE}}=-1$ (cosmological constant). 
Legend is the same as Fig.~\ref{fig-1}. 
}
\label{fig-2}
\end{figure}
\end{center}

\begin{center}
\begin{figure}[tbp]
\resizebox{!}{6.5cm}{
  \includegraphics{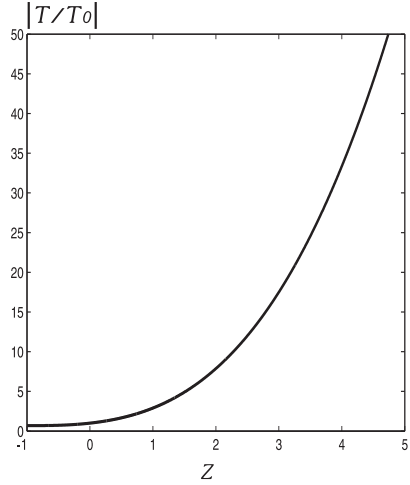}
                  }
\caption{$\abs{ T/T_0 }$ 
as a function of the redshift $z$ for $\abs{p}=0.1$.
}
\label{fig-3}
\end{figure}
\end{center}
%

\begin{center}
\begin{figure}[tbp]
\begin{tabular}{ll}
\begin{minipage}{80mm}
\begin{center}
\unitlength=1mm
\resizebox{!}{6.5cm}{
   \includegraphics{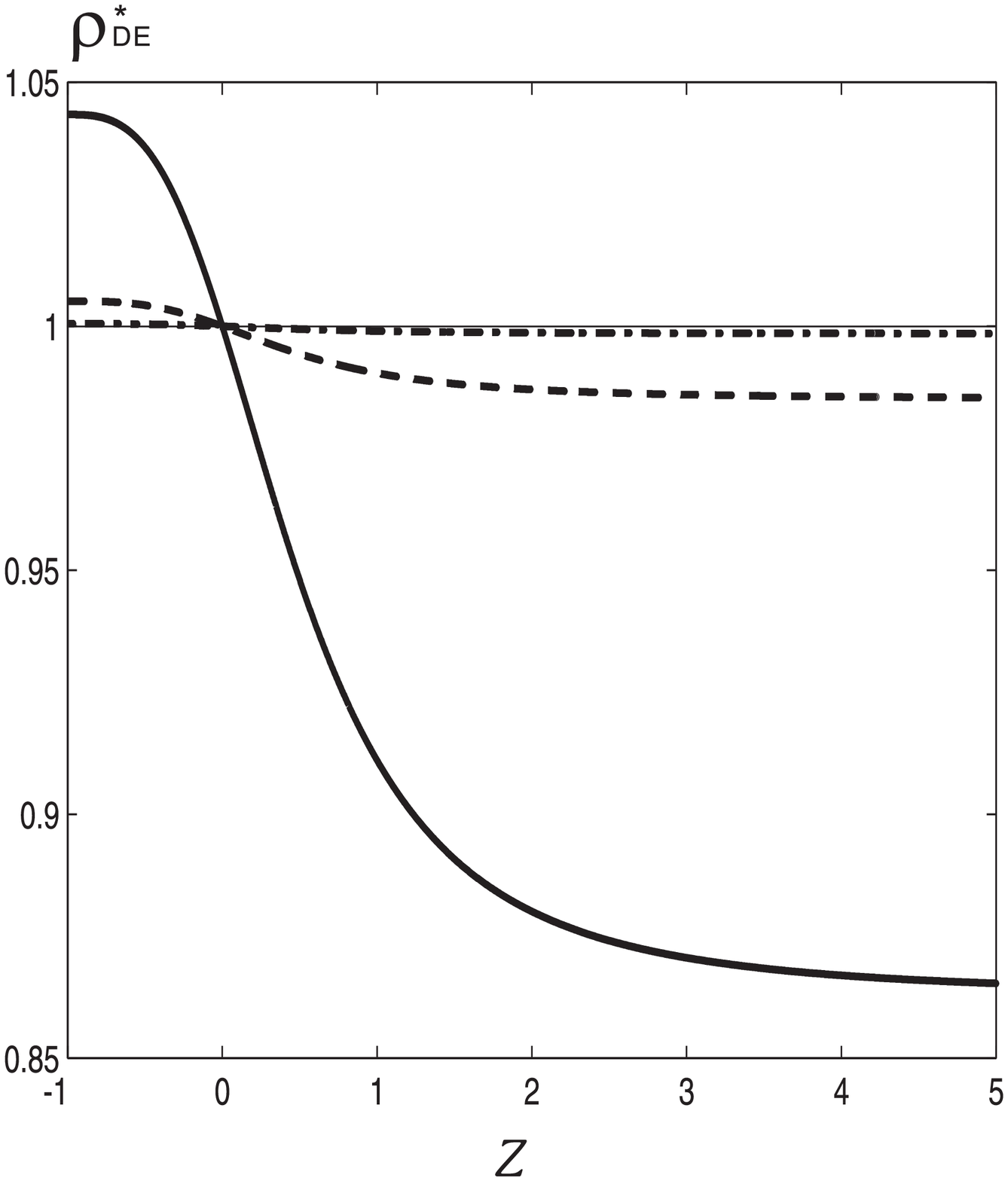}
                  }
\end{center}
\end{minipage}
&
\begin{minipage}{80mm}
\begin{center}
\unitlength=1mm
\resizebox{!}{6.5cm}{
   \includegraphics{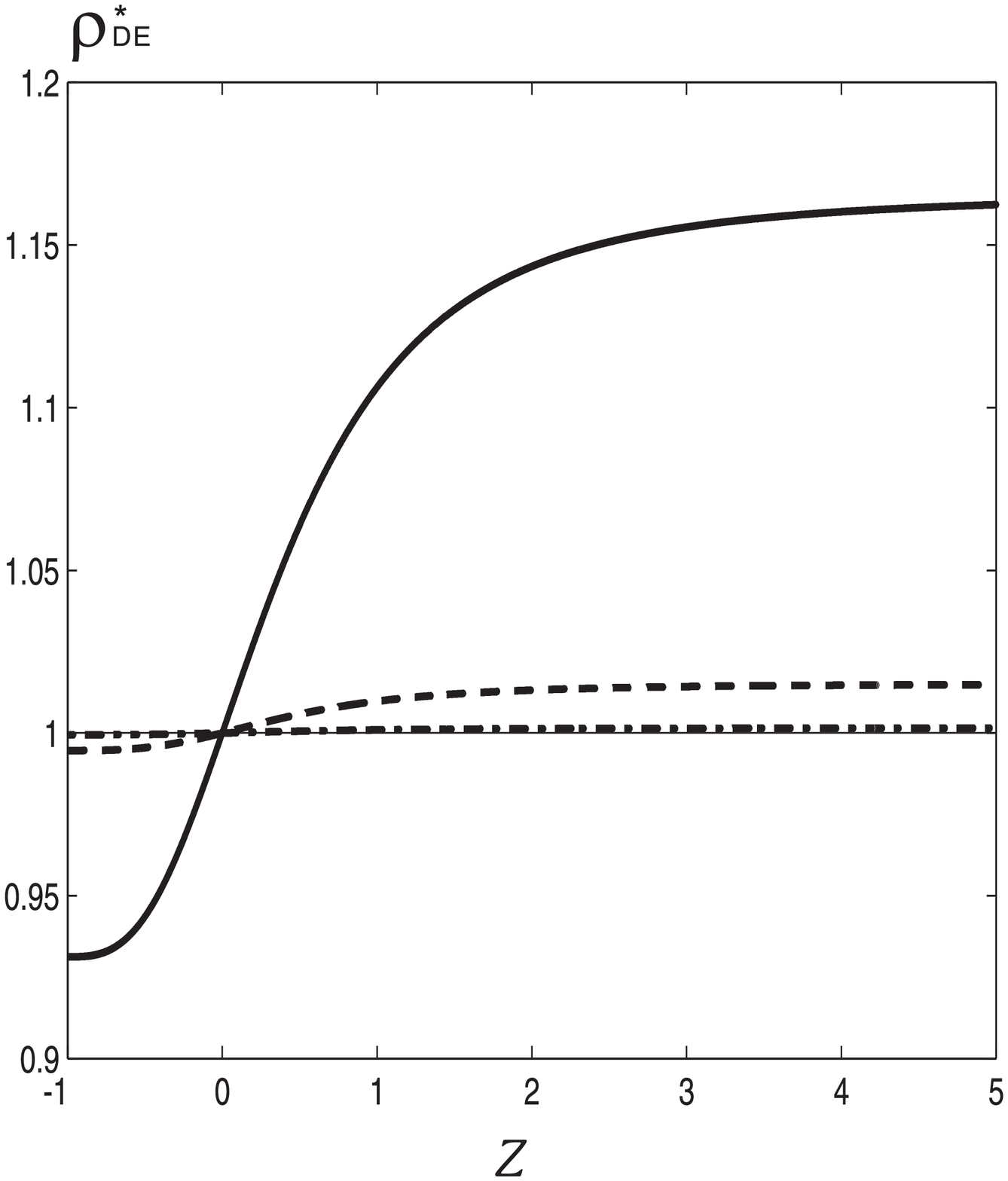}
                  }
\end{center}
\end{minipage}

\end{tabular}
\caption{$\rho^{\star}_{\mathrm{DE}} \equiv 
\rho_{\mathrm{DE}}/\rho_{\mathrm{DE}}^{(0)}$ 
as a function of the redshift $z$, 
where the thin solid line shows $w_{\mathrm{DE}}=-1$ (cosmological constant). 
Legend is the same as Fig.~\ref{fig-1}. 
}
\label{fig-4}
\end{figure}
\end{center}


We display $w_{\mathrm{DE}}$ 
as a function of $\abs{ T/T_0 }$ in Fig.~\ref{fig-2},
where
 the left and right panels are for $p>0$ and $p<0$, respectively.
 In Fig.~\ref{fig-3},
we show the cosmological evolution of 
$\abs{ T/T_0 }$ as a function of the redshift $z$ for $\abs{p}=0.1$.
The qualitative behaviors of $\abs{ T/T_0 }$ 
for $\abs{p}=0.01$ and $0.001$ 
are similar to those for $\abs{p}=0.1$. 
In addition, 
the cosmological evolution of $\rho^{\star}_{\mathrm{DE}} \equiv 
\rho_{\mathrm{DE}}/\rho_{\mathrm{DE}}^{(0)}$ 
as a function of the redshift $z$ 
is shown in Fig.~\ref{fig-4}. 

{}From Fig.~\ref{fig-2}, we find that the crossing of the phantom divide 
occurs around $\abs{ T/T_0 } \approx 0.7 < 1$. 
This means that 
the crossing 
cannot be realized in the past 
and far future,
unlike those in $f(R)$ models~\cite{B-G-L}. 
Explicitly, by denoting $\abs{ T/T_0 }_{\mathrm{cross}}$ being the crossing 
point,
we have $\abs{ T/T_0 }_{\mathrm{cross}}
\simeq
 0.772$,  $0.744$,  $0.740$, $0.689$, $0.736$, and $0.740$
for 
$p=0.1$, $0.01$, $0.001$, 
$-0.1$, $-0.01$, and $-0.001$, 
respectively. 
Clearly, 
the universe reaches
the line $w_{\mathrm{DE}}=-1$ 
without a crossing even in the very far future ($z \simeq -1$). 

In the high $z$ regime, the universe is at the matter-dominated stage 
and therefore, 
$\abs{ T }=6H^2= 16 \pi G \left(\rho_{\mathrm{m}}+\rho_{\mathrm{DE}}\right)$ 
decreases with time. 
Around the present time ($z=0$), 
$\rho_{\mathrm{m}}$ is smaller than $\rho_{\mathrm{DE}}$ 
and dark energy becomes dominant over matter. 
Fig.~\ref{fig-3} illustrates that 
$\abs{ T/T_0 }$ decreases monotonously with $z$. 
At the infinite future, $\abs{ T/T_0 }_{\mathrm{cross}}$ is reached.
However, the the universe is in either the phantom phase ($p>0$) or non-phantom phase ($p<0$) without the crossing.
Since by that time the contribution from non-relativistic matter and 
radiation is supposed to be negligible, such an evolution pattern of 
$\abs{ T/T_0 }$ is driven by the $f(T)$ term. 
In the future ($z<0$), 
$\rho_{\mathrm{DE}}$ arrives at the crossing point and finally 
becomes constant as shown in Fig.~\ref{fig-4}. 

The right-hand side of the first equality in Eq.~(\ref{eq:2.14}) implies 
that when $w_{\mathrm{DE}}$ crosses the phantom divide line of 
$w_{\mathrm{DE}} = -1$, 
the term
$T^\prime \left( f_T+2Tf_{TT} \right)/\left[3T 
\left( f/T-2f_T \right) \right]$ 
has to flip the sign.
Since  the sign of $T$ is fixed to be negative due to
$T=-6H^2$,
the sufficient and necessary conditions for 
the crossing of the phantom divide is 
that the sign of the combination 
$T^\prime \left( f_T+2Tf_{TT} \right)/
\left( f/T-2f_T \right)$ 
changes in the cosmological evolution. 
Note that 
the sign of $\dot{H}$ is opposite to $T^\prime$ since
$T^\prime = -12 H H^\prime = -12 \dot{H}$.
When the energy density of dark energy becomes perfectly dominant over 
those of non-relativistic matter and radiation, 
one has $w_{\mathrm{DE}} \approx 
w_{\mathrm{eff}} \equiv 
-1 -2\dot{H}/\left(3H^2\right) 
= P_{\mathrm{tot}}/\rho_{\mathrm{tot}} 
$, 
where $w_{\mathrm{eff}}$ is the effective equation of state 
for the universe~\cite{Review-N-O}, 
and 
$\rho_{\mathrm{tot}} \equiv \rho_{\mathrm{DE}} + \rho_{\mathrm{m}} 
+ \rho_{\mathrm{r}}
$ 
and 
$P_{\mathrm{tot}} \equiv P_{\mathrm{DE}} 
+ P_{\mathrm{r}}
$ 
are the total energy density and pressure of the universe, 
with $\rho_{\mathrm{r}}$ and $P_{\mathrm{r}}$ being
the energy density and pressure of radiation, 
respectively. 
For $\dot{H} < 0\ (>0)$, $w_\mathrm{eff} >-1\ (<-1)$, representing the non-phantom (phantom)
phase, 
while 
$w_\mathrm{eff} =-1$ for  $\dot{H} = 0$, corresponding to the 
cosmological constant. 
Clearly, the physical 
reason why the crossing of the phantom divide appears 
is that the sign of $\dot{H}$ 
changes from negative (the non-phantom phase) to 
positive (the phantom phase) 
due to the dominance of dark energy over 
non-relativistic matter and radiation. 

We now investigate 
why the crossing of the phantom divide cannot occur 
in the exponential $f(T)$ theory in Eq.~(\ref{eq:3.4}) for $p\leq 1$. 
{}From Eq.~(\ref{eq:3.4}), we obtain
\begin{eqnarray}
f_T \Eqn{=} 
\alpha \left( 1-e^{pT_0/T}+\frac{pT_0}{T} 
e^{pT_0/T}\right)\,, 
\label{eq:3.E1-01} \\ 
f_{TT} \Eqn{=} 
-\alpha \left(\frac{pT_0}{T}\right)^2 
\frac{1}{T} e^{pT_0/T}\,.
\label{eq:3.E1-02}
\end{eqnarray}
%
To illustrate our results,
we only concentrate on the limits of $0< p \ll 1$ and 
$X \equiv pT_0/T \ll 1$
In this case, $T_0/T \lesssim 1$, which corresponds to the region 
from the far past to the near future. 
Consequently,
Eqs.~(\ref{eq:3.4}), (\ref{eq:3.E1-01}) and (\ref{eq:3.E1-02}) are 
approximately expressed as  
\begin{equation}
\frac{f}{T} \approx -\alpha \left(X+ \frac{X^2}{2}\right)\,, 
\quad 
f_T \approx \frac{\alpha X^2}{2}\,, 
\quad 
Tf_{TT} \approx -\alpha X^2\,.
\label{eq:3.E1-03}
\end{equation}
By using these equations, we find 
%
\begin{eqnarray}
\frac{f}{T}-2f_T 
\Eqn{\approx} -\alpha X \left(1+ \frac{3X}{2}\right)\,, 
\label{eq:3.E1-04} \\
f_T+2Tf_{TT} \Eqn{\approx} -\frac{3\alpha X^2}{2}\,. 
\label{eq:3.E1-05}
\end{eqnarray}
{}From Eqs.~(\ref{eq:3.E1-04}) and (\ref{eq:3.E1-05}), we see that 
the signs of both $f/T-2f_T$ and $f_T+2Tf_{TT}$ do not change. 
Note that 
the sign of $T^\prime$ is also unchanged 
because $\abs{ T/T_0 }$ decreases monotonously with $z$ until 
$\abs{ T/T_0 }_{\mathrm{min}}$ as shown in Fig.~\ref{fig-3}. 
As a result, the crossing of the phantom divide cannot be realized 
in the exponential $f(T)$ theory. 

\begin{center}
\begin{figure}[tbp]
\begin{tabular}{ll}
\begin{minipage}{80mm}
\begin{center}
\unitlength=1mm
\resizebox{!}{6.5cm}{
   \includegraphics{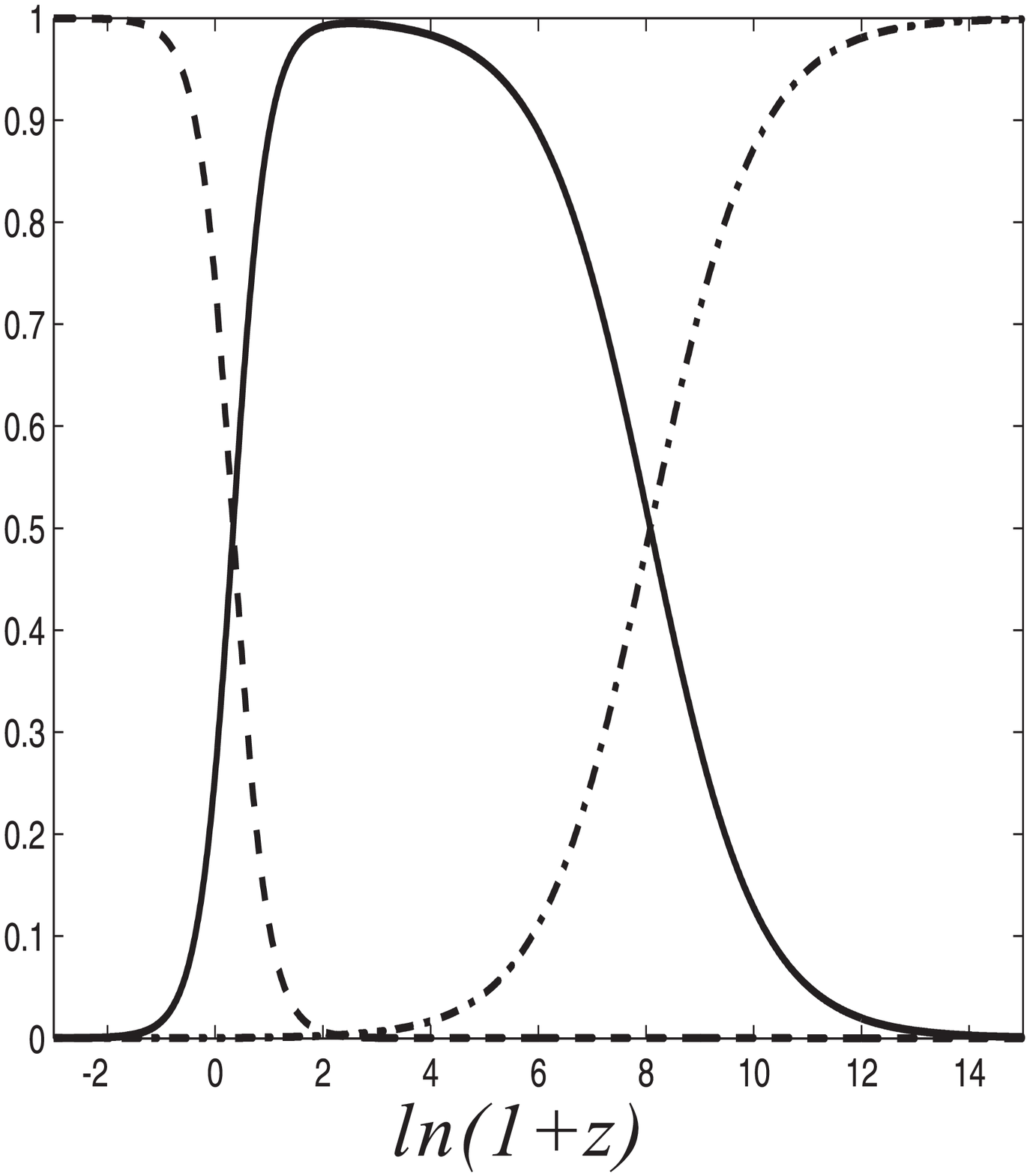}
                  }
\end{center}
\end{minipage}
&
\begin{minipage}{80mm}
\begin{center}
\unitlength=1mm
\resizebox{!}{6.5cm}{
   \includegraphics{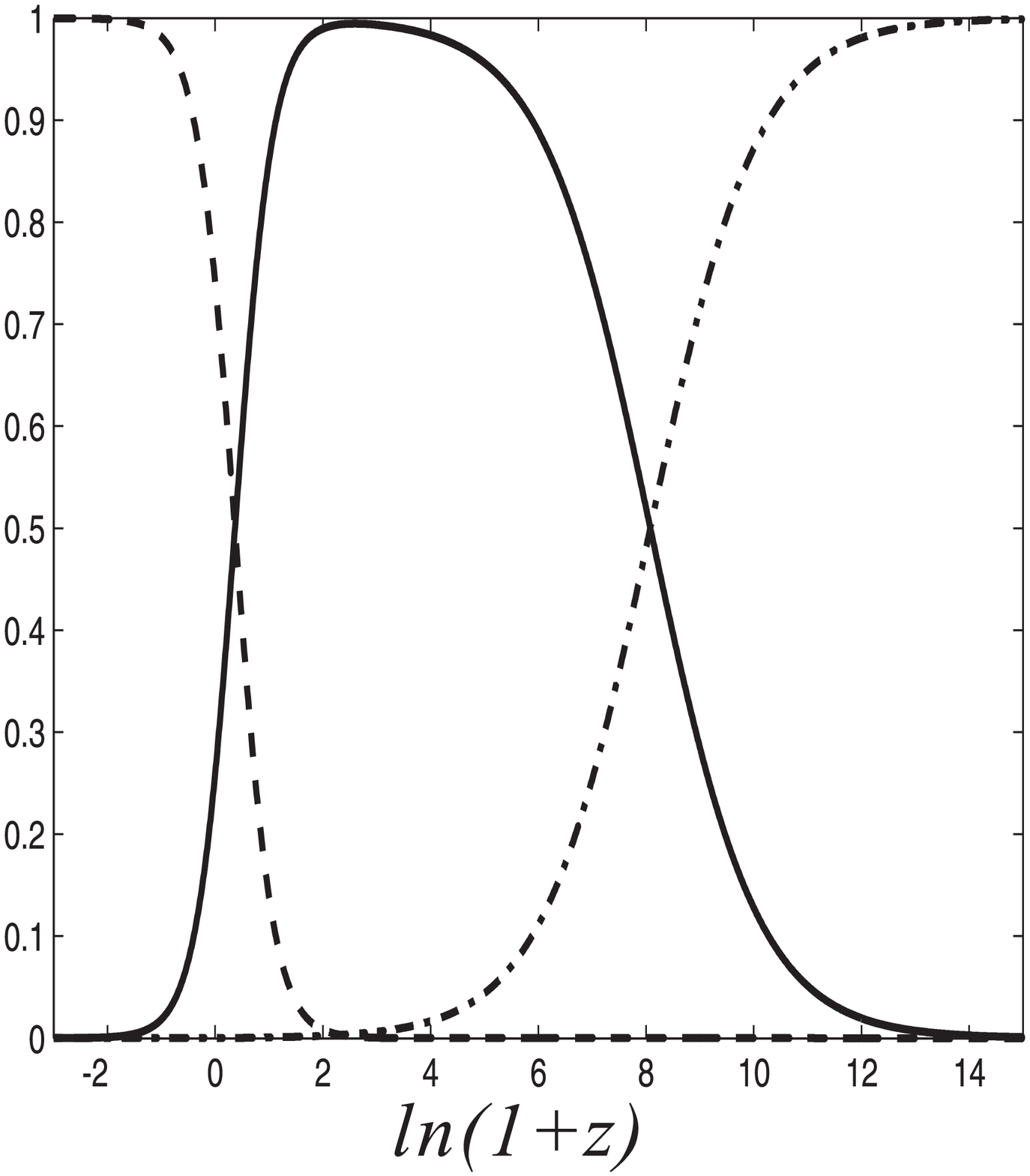}
                  }
\end{center}
\end{minipage}

\end{tabular}
\caption{$\Omega_{\mathrm{DE}}$ (dashed line), 
$\Omega_{\mathrm{m}}$ (solid line) and 
$\Omega_{\mathrm{r}}$ (dash-dotted line) 
as functions of the redshift $z$ 
for $\abs{p}=0.1$. 
Legend is the same as Fig.~\ref{fig-1}. 
}
\label{fig-5}
\end{figure}
\end{center}


In Fig.~\ref{fig-5}, 
we illustrate the fractional densities of dark energy 
($\Omega_{\mathrm{DE}} \equiv \rho_{\mathrm{DE}}/\rho_{\mathrm{crit}}^{(0)}$), 
non-relativistic matter 
($\Omega_{\mathrm{m}} \equiv \rho_{\mathrm{m}}/\rho_{\mathrm{crit}}^{(0)}$) 
and radiation ($\Omega_{\mathrm{r}} \equiv 
\rho_{\mathrm{r}}/\rho_{\mathrm{crit}}^{(0)}$) 
as functions of the redshift $z$ 
for $\abs{p}=0.1$. 
The cosmological evolutions of $\Omega_{\mathrm{DE}}$, $\Omega_{\mathrm{m}}$ 
and $\Omega_{\mathrm{r}}$ for $\abs{p}=0.01$ and $0.001$ 
are similar to those for $\abs{p}=0.1$. 
In order to analyze 
not only non-relativistic matter 
and dark energy but also radiation, 
we have included the contribution from radiation in Eq.~(\ref{eq:3.1}) 
and used the following variable~\cite{Hu:2007nk}: 
\begin{equation}
y_H 
= \frac{H^2}{\bar{m}^2} -a^{-3} -\chi a^{-4}\,, 
\label{eq:3.6}
\end{equation}
with
\begin{equation}
\chi \equiv \frac{\rho_{\mathrm{r}}^{(0)}}{\rho_{\mathrm{m}}^{(0)}}
\simeq 3.1 \times 10^{-4}\,, 
\label{eq:3.7}
\end{equation}
where 
$\rho_{\mathrm{r}}^{(0)}$ is the energy density of radiation 
at the present time. 

In the high $z$ regime ($z \gtrsim 3225$), the universe is at the 
radiation-dominated stage ($\Omega_{\mathrm{r}} \gg \Omega_{\mathrm{DE}}$, 
$\Omega_{\mathrm{r}} > \Omega_{\mathrm{m}}$). 
As $z$ decreases, 
the universe enters the matter-dominated stage 
($\Omega_{\mathrm{m}} > \Omega_{\mathrm{DE}}$, 
$\Omega_{\mathrm{m}} \gg \Omega_{\mathrm{r}}$). 
After that, 
eventually dark energy becomes 
dominant over matter for $z<z_{\mathrm{DE}}$, 
where $z_{\mathrm{DE}}$ is the crossover point in which 
$\Omega_{\mathrm{DE}} = \Omega_{\mathrm{m}}$. 
Explicitly, we have $z_{\mathrm{DE}} = 0.40$ and $0.44$ for 
$p=0.1$ and $-0.1$, respectively. 
At the present time ($z=0$), as an initial condition we have used 
$(\Omega_{\mathrm{DE}}^{(0)}, \Omega_{\mathrm{m}}^{(0)}, 
\Omega_{\mathrm{r}}^{(0)}) = 
(0.74, 0.26, 8.1 \times 10^{-5})$. 
Note that $z_{\mathrm{DE}} = 0.42$ 
in the $\Lambda\mathrm{CDM}$ model. 
As a result, the current accelerated expansion of the universe following 
the radiation-dominated and matter-dominated stages can be achieved in 
the exponential $f(T)$ theory.

\subsection{Logarithmic $f(T)$ theory}

In this subsection, 
we examine a logarithmic $f(T)$ theory, given by 
\begin{equation}
f(T)= \beta T_0 \left(\frac{qT_0}{T}\right)^{-1/2} \ln 
\left(\frac{qT_0}{T}\right)\,,
\label{eq:4.1} 
\end{equation}
with
\begin{equation}
\beta \equiv \frac{1-\Omega_{\mathrm{m}}^{(0)}}{2q^{-1/2}}\,,
\label{eq:4.2} 
\end{equation}
where $q$ is a positive constant. 
We note that the theory in Eq.~(\ref{eq:4.1}) contains only one parameter 
$q$ if the value of $\Omega_{\mathrm{m}}^{(0)}$ is obtained 
in the same way as the exponential $f(T)$ theory. 

\begin{center}
\begin{figure}[tbp]
\begin{tabular}{ll}
\begin{minipage}{80mm}
\begin{center}
\unitlength=1mm
\resizebox{!}{6.5cm}{
   \includegraphics{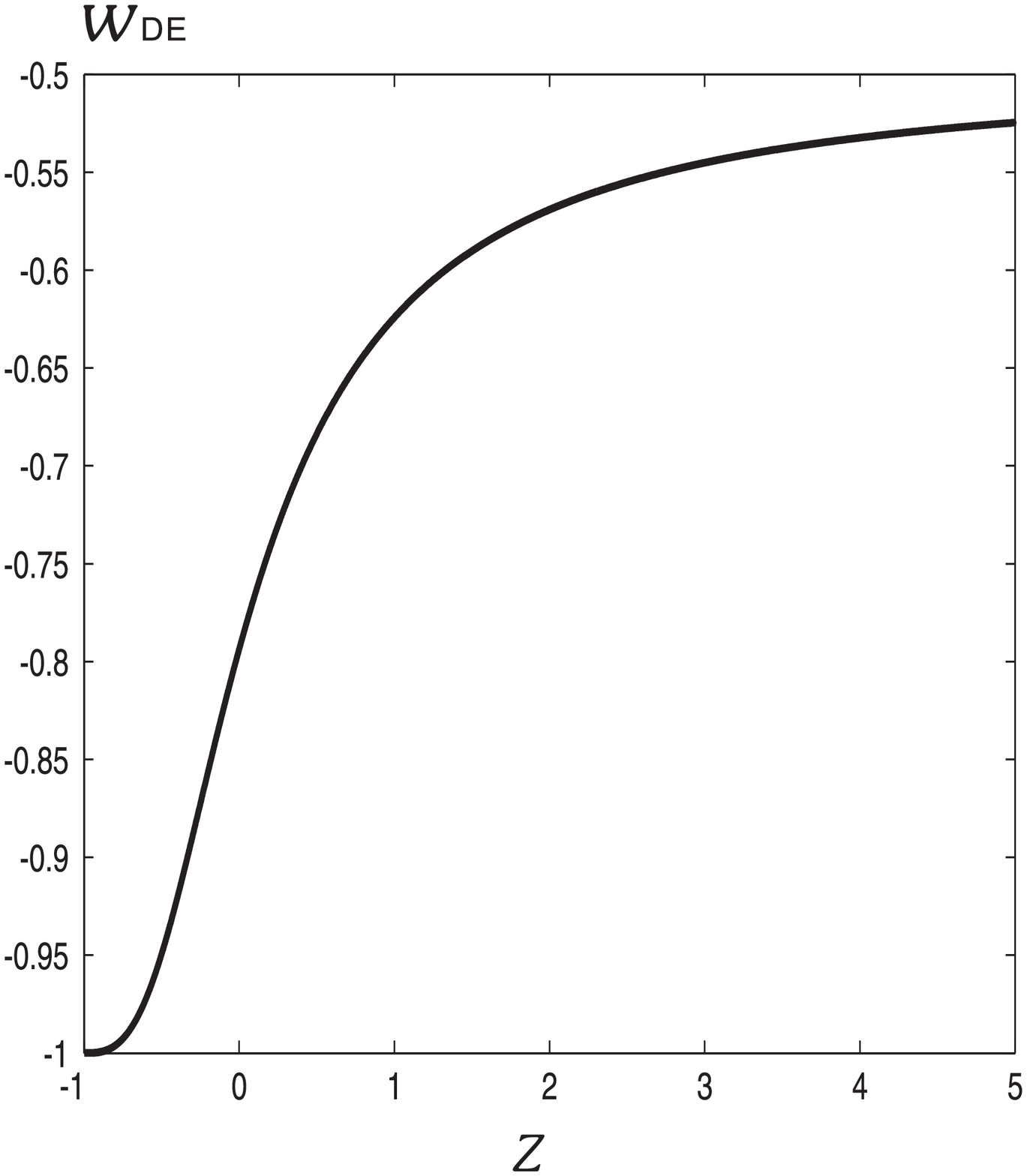}
                  }
\end{center}
\end{minipage}
&
\begin{minipage}{80mm}
\begin{center}
\unitlength=1mm
\resizebox{!}{6.5cm}{
   \includegraphics{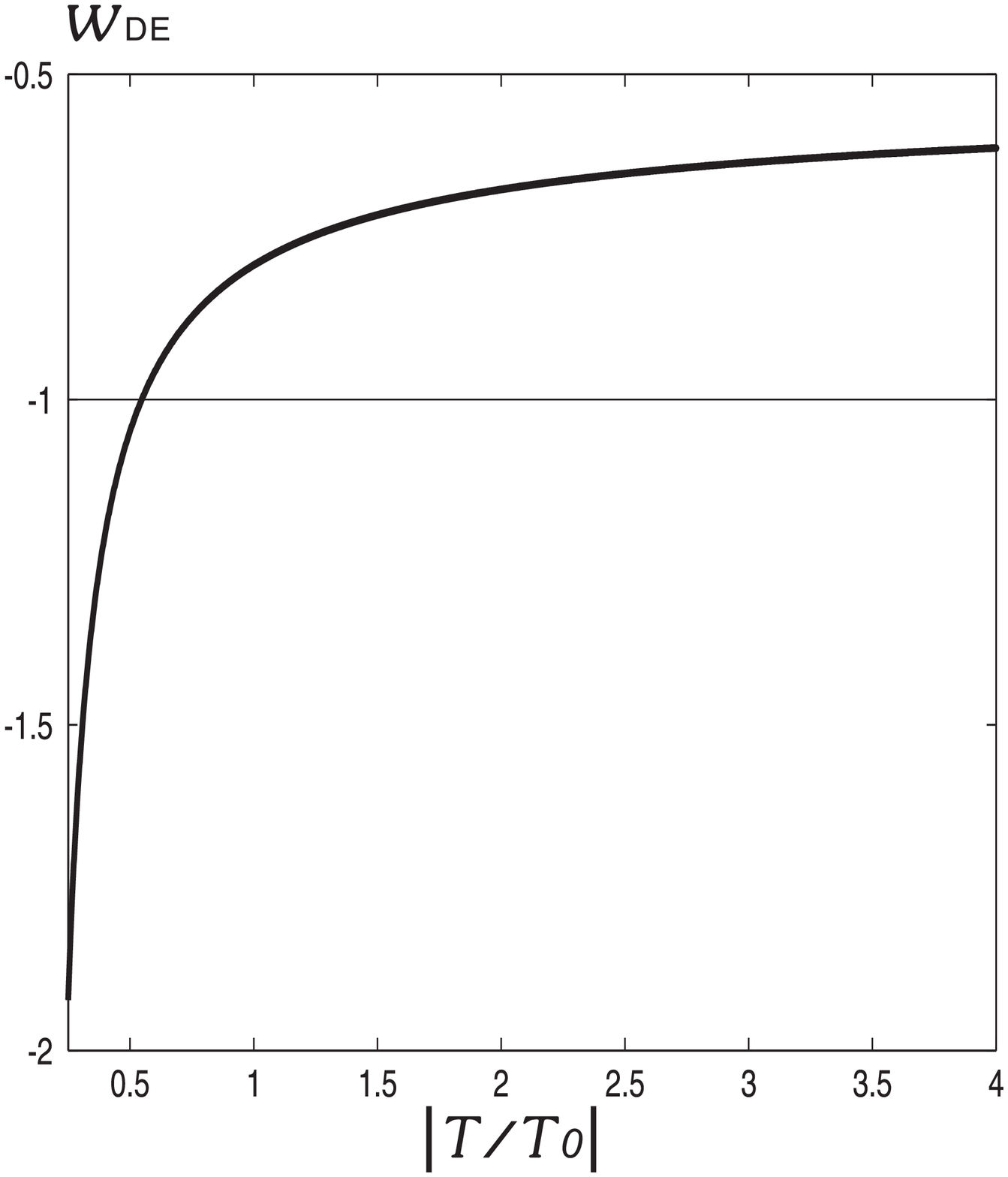}
                  }
\end{center}
\end{minipage}

\end{tabular}
\caption{$w_{\mathrm{DE}}$ 
as functions of $z$ (left panel) 
and $\abs{ T/T_0 }$ (right panel) for $q=1$ and 
$\Omega_{\mathrm{m}}^{(0)} = 0.26$ in 
the logarithmic $f(T)$ theory, 
where the thin solid line shows $w_{\mathrm{DE}}=-1$ (cosmological constant). 
}
\label{fig-6}
\end{figure}
\end{center}

In Fig.~\ref{fig-6}, we draw $w_{\mathrm{DE}}$ 
as functions of $z$ (left panel) 
and $\abs{ T/T_0 }$ (right panel) for $q=1$ and 
$\Omega_{\mathrm{m}}^{(0)} = 0.26$ in 
the logarithmic $f(T)$ theory 
in Eq.~(\ref{eq:4.1}). 
{}From Fig.~\ref{fig-6}, we observe that $w_{\mathrm{DE}}$ does not 
cross the phantom divide line $w_{\mathrm{DE}}=-1$, 
similar to the exponential theory. 
In this logarithmic theory, 
the universe is always in the non-phantom phase 
($w_{\mathrm{DE}} > -1$) 
and 
$\abs{ T/T_0 }$ decreases monotonously with $z$. 
The present value of $w_{\mathrm{DE}}$ is 
$w_{\mathrm{DE}} (z=0) = -0.79$ for $q=1$.  
The right figure in Fig.~\ref{fig-6} indicates that the crossing of the 
phantom divide 
occurs at $\abs{ T/T_0 }_{\mathrm{cross}} = 0.547600$ 
and 
the minimum of $\abs{ T/T_0 }$ 
is $\abs{ T/T_0 }_{\mathrm{min}} = 0.547600$. 
Thus, 
the universe does not 
cross the line of $w_{\mathrm{DE}}=-1$ even in the very far future 
($z \simeq -1$). 

We remark that 
$w_{\mathrm{DE}}$ is qualitatively 
independent of 
the value of $q$. 
This is because 
in the logarithmic $f(T)$ theory, 
$w_{\mathrm{DE}}$ and $\rho_{\mathrm{DE}}$ can be expressed as 
\begin{equation}
w_{\mathrm{DE}} 
= -\frac{1}{2-\left(1-\Omega_{\mathrm{m}}^{(0)}\right)
\left(T/T_0\right)^{-1/2}}
\label{eq:4.a1} 
\end{equation} 
and 
\begin{equation}
\rho_{\mathrm{DE}} 
= -\frac{T}{16 \pi G} 
\left(1-\Omega_{\mathrm{m}}^{(0)}\right) 
\left(\frac{T}{T_0}\right)^{-1/2}\,,
\label{eq:4.a2}
\end{equation}
respectively. 
Similar to the exponential $f(T)$ theory,
%
%
%
%
it is easy to show that the crossing of the phantom divide cannot occur 
in the logarithmic $f(T)$ theory. 

\section{Realizing the crossing of the phantom divide in $f(T)$ theory} 

As shown in the previous section, 
the universe always stays in the phantom and non-phantom phases 
in the exponential ($p>0$) and logarithmic $f(T)$ theories, respectively. 
In this section, we investigate the cosmological evolution in 
a combined $f(T)$ theory with both logarithmic and exponential terms 
and check if the crossing of the phantom divide can happen 
in this combined theory. 
The explicit form of the theory is given by 
\begin{equation}
f(T)= \gamma \left[ T_0 \left(\frac{uT_0}{T}\right)^{-1/2} \ln 
\left(\frac{uT_0}{T}\right) 
-T \left(1-e^{uT_0/T}\right) 
\right] 
\label{eq:4.3} 
\end{equation}
with
\begin{equation}
\gamma \equiv \frac{1-\Omega_{\mathrm{m}}^{(0)}}{2u^{-1/2}
+\left[1-\left(1-2u\right)e^u \right]}\,,
\label{eq:4.4} 
\end{equation}
where $u$ is a constant. 
Here, we have taken $p=q=u >0$ to simplify our discussions 
when we combine Eqs.~(\ref{eq:3.4}) and (\ref{eq:4.1}). 
We note that the model in Eq.~(\ref{eq:4.3}) contains only one parameter 
$u$ if one has the value of $\Omega_{\mathrm{m}}^{(0)}$ 
like the exponential and logarithmic models. 

\begin{center}
\begin{figure}[tbp]
\resizebox{!}{6.5cm}{
   \includegraphics{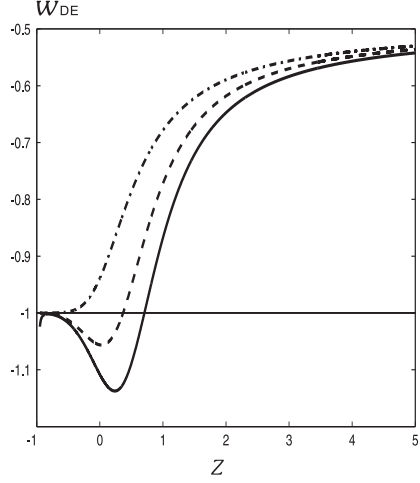}
                  }
\caption{$w_{\mathrm{DE}}$ 
as a function of the redshift $z$ for $u=1$ (solid line), 
$0.8$ (dashed line), $0.5$ (dash-dotted line) and 
$\Omega_{\mathrm{m}}^{(0)} = 0.26$ in 
the combined $f(T)$ theory with both 
logarithmic and exponential terms 
in Eq.~(\ref{eq:4.3}), 
where the thin solid line shows $w_{\mathrm{DE}}=-1$ (cosmological constant). 
}
\label{fig-7}
\end{figure}
\end{center}

\begin{center}
\begin{figure}[tbp]
\resizebox{!}{6.5cm}{
   \includegraphics{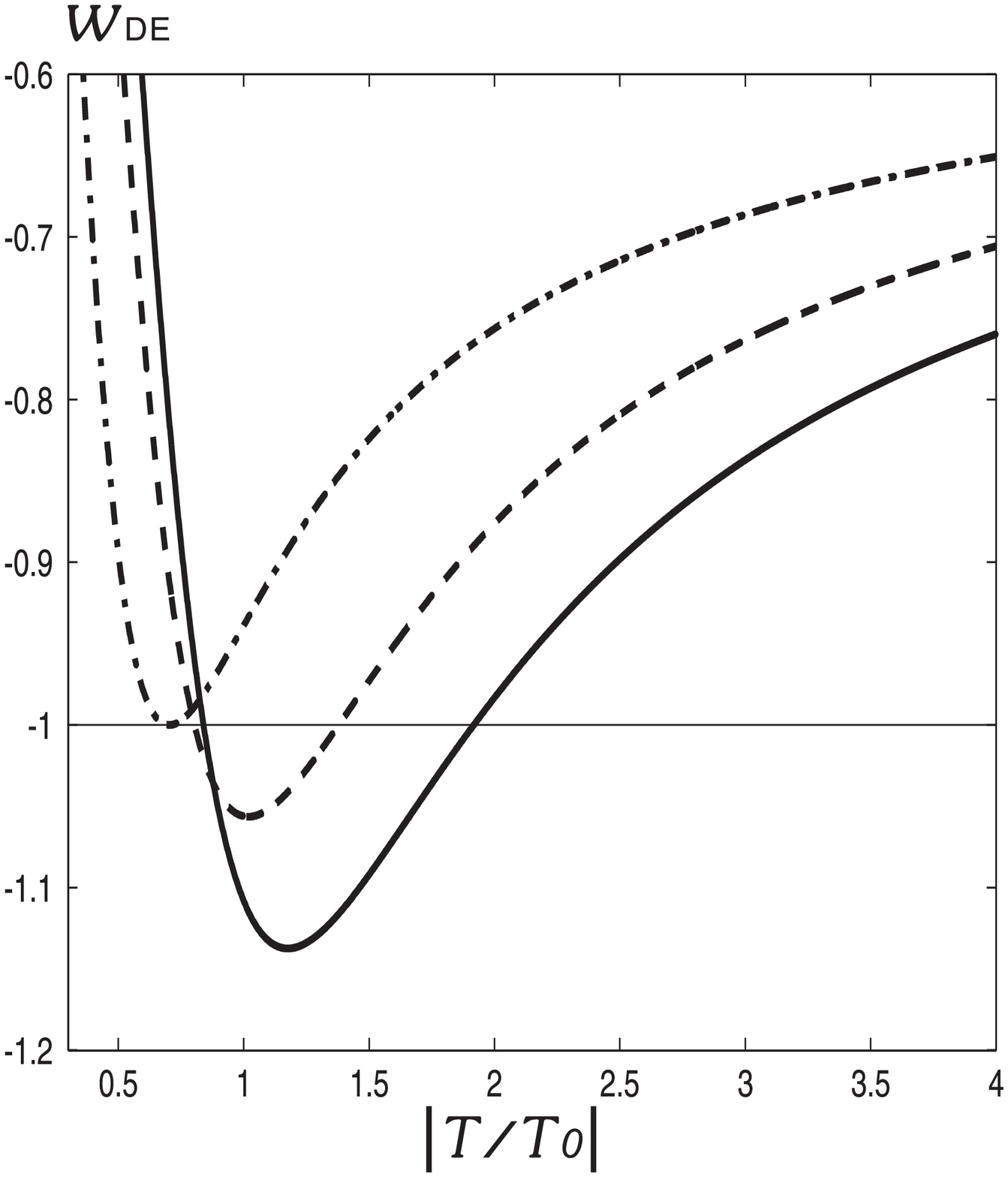}
                  }
\caption{$w_{\mathrm{DE}}$ 
as a function of $\abs{ T/T_0 }$. 
Legend is the same as Fig.~\ref{fig-7}. 
}
\label{fig-8}
\end{figure}
\end{center}

\begin{center}
\begin{figure}[tbp]
\resizebox{!}{6.5cm}{
   \includegraphics{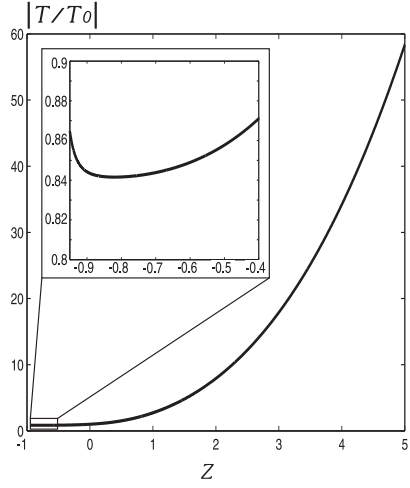}
                  }
\caption{$\abs{ T/T_0 }$ 
as a function of the redshift $z$ for $u=1$. 
The range of $-0.95 \leq z \leq -0.40$ is magnified 
in the small window. 
Legend is the same as Fig.~\ref{fig-7}. 
}
\label{fig-9}
\end{figure}
\end{center}

\begin{center}
\begin{figure}[tbp]
\resizebox{!}{6.5cm}{
   \includegraphics{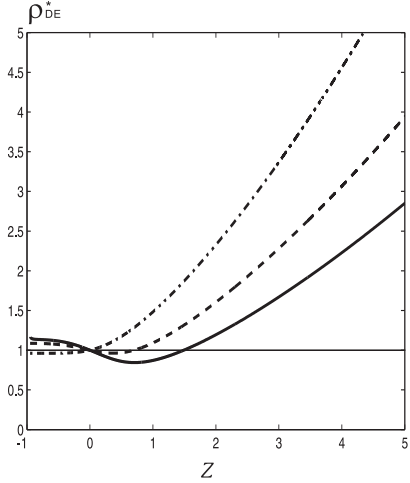}
                  }
\caption{$\rho^{\star}_{\mathrm{DE}} \equiv 
\rho_{\mathrm{DE}}/\rho_{\mathrm{DE}}^{(0)}$ 
as a function of the redshift $z$. 
Legend is the same as Fig.~\ref{fig-7}. 
}
\label{fig-10}
\end{figure}
\end{center}

In Fig.~\ref{fig-7}, we plot the equation of state for dark energy 
$w_{\mathrm{DE}}$ 
as a function of the redshift $z$ for $u=1$, $0.8$ and $0.5$ 
and $\Omega_{\mathrm{m}}^{(0)} = 0.26$ 
in the combined $f(T)$ theory in Eq.~(\ref{eq:4.3}). 
As seen from Fig.~\ref{fig-7}, it is clear that $w_{\mathrm{DE}}$ can 
cross $w_{\mathrm{DE}}=-1$ 
at $z = z_{\mathrm{cross}}$. 
The universe 
evolves from the non-phantom phase ($w_{\mathrm{DE}} > -1$) to 
the phantom one ($w_{\mathrm{DE}} < -1$) and 
finally approaches $w_{\mathrm{DE}}=-1$. 
The present values of $w_{\mathrm{DE}}$ are 
$w_{\mathrm{DE}} (z=0) = -1.1$, $-1.1$ and $-0.94$ for 
$u=1$, $0.8$ and $0.5$, respectively. 
The crossing values of $z_{\mathrm{cross}}$ are 
$0.70$ and $0.36$ for 
$u=1$ and $0.8$, respectively. 
It is interesting to note that 
such a crossing behavior is consistent with the recent cosmological 
observational data~\cite{observational status}, 
but it is opposite to those in the viable $f(R)$ 
theories~\cite{B-G-L}. 
We note that 
for $u \leq 0.5$, the universe asymptotically approaches 
$w_{\mathrm{DE}}=-1$ 
without crossing it 
even in the very far future. 
The cosmological evolution of $w_{\mathrm{DE}}$ 
as a function of $\abs{ T/T_0 }$ is shown 
in Fig.~\ref{fig-8}, 
while that of $\abs{ T/T_0 }$ as a function of $z$ for $u=1$ 
is depicted in Fig.~\ref{fig-9}. 
In addition, 
$\rho^{\star}_{\mathrm{DE}} \equiv 
\rho_{\mathrm{DE}}/\rho_{\mathrm{DE}}^{(0)}$ 
as a function of $z$ 
is given in Fig.~\ref{fig-10}. 

We now demonstrate that
the crossing of the phantom divide can occur 
in the combined $f(T)$ theory in Eq.~(\ref{eq:4.3}).
In this theory, both signs of $T$ and $T^\prime$ are unchanged like the exponential 
$f(T)$ theory.
{}From Eq.~(\ref{eq:4.3}), we obtain
\begin{eqnarray}
f_T \Eqn{=} -\frac{\gamma}{u} \sqrt{\frac{uT_0}{T}} 
\left(1-\frac{1}{2} \ln \left(\frac{uT_0}{T}\right) \right)
- \gamma \left( 1-e^{uT_0/T}+\frac{uT_0}{T} 
e^{uT_0/T}\right)\,, 
\nonumber \\ 
f_{TT} \Eqn{=} 
-\frac{\gamma}{4u^2 T_0} \left(\frac{uT_0}{T}\right)^{3/2} 
\ln \left(\frac{uT_0}{T}\right)
+ \frac{\gamma}{u T_0} \left(\frac{uT_0}{T}\right)^{3} 
e^{uT_0/T}\,,
\label{eq:4.E3-02}
\end{eqnarray}
%
%
leading to
\begin{eqnarray}
\frac{f}{T}-2f_T \Eqn{\approx} 
\gamma \sqrt{\frac{uT_0}{T}} \left[ 
\frac{2}{u} + \sqrt{\frac{uT_0}{T}} 
\left(1+ \frac{3}{2} \frac{uT_0}{T} \right)\right]\,, 
\nonumber\\
f_T+2Tf_{TT} \Eqn{\approx}  
-\gamma \sqrt{\frac{uT_0}{T}} \left[ 
\frac{1}{u} - \frac{3}{2} \left( \frac{uT_0}{T} \right)^{3/2}
\right]\,, 
\label{eq:4.E3-04}
\end{eqnarray}
where 
we have only concentrated on the case of $uT_0/T < 1$ 
to simplify our discussion.
{}From 
Eq.~(\ref{eq:4.E3-04}), we see that 
the sign of $f/T-2f_T$ is fixed, while
that of $f_T+2Tf_{TT}$ can change when 
$uT_0/T$ becomes larger or smaller than the value of 
$uT_0/T = \left[2/\left(3u\right)\right]^{2/3}$. 
Hence, 
the sufficient and necessary conditions for 
the crossing of the phantom divide in the combined $f(T)$ theory can be satisfied. 

\begin{center}
\begin{figure}[tbp]
\resizebox{!}{6.5cm}{
   \includegraphics{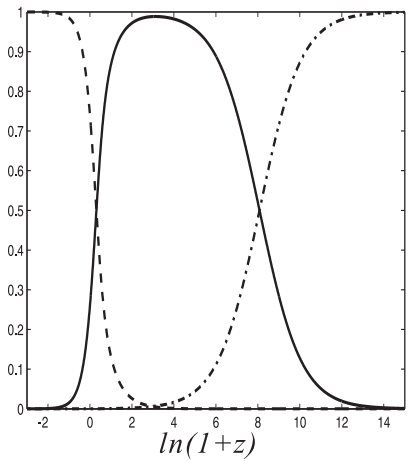}
                  }
\caption{$\Omega_{\mathrm{DE}}$ (dashed line), 
$\Omega_{\mathrm{m}}$ (solid line) and 
$\Omega_{\mathrm{r}}$ (dash-dotted line) 
as functions of the redshift $z$ 
for $u=1$. 
Legend is the same as Fig.~\ref{fig-7}. 
}
\label{fig-11}
\end{figure}
\end{center}


In Fig.~\ref{fig-11}, 
we demonstrate the fractional densities of dark energy 
($\Omega_{\mathrm{DE}} \equiv \rho_{\mathrm{DE}}/\rho_{\mathrm{crit}}^{(0)}$), 
non-relativistic matter 
($\Omega_{\mathrm{m}} \equiv \rho_{\mathrm{m}}/\rho_{\mathrm{crit}}^{(0)}$) 
and radiation ($\Omega_{\mathrm{r}} \equiv 
\rho_{\mathrm{r}}/\rho_{\mathrm{crit}}^{(0)}$) 
as functions of $z$ 
for $u=1$. 
The evolutions 
of $\abs{ T/T_0 }$, 
$\Omega_{\mathrm{DE}}$, $\Omega_{\mathrm{m}}$ 
and $\Omega_{\mathrm{r}}$ for $u=0.8$ and $u=0.5$ 
are similar to those for $u=1$. 
In the high $z$ regime ($z \gtrsim 3225$), the universe is at the 
radiation-dominated stage 
($\Omega_{\mathrm{r}} \gg \Omega_{\mathrm{DE}}$, 
$\Omega_{\mathrm{r}} > \Omega_{\mathrm{m}}$). 
As $z$ decreases, 
the matter-dominated stage 
($\Omega_{\mathrm{m}} > \Omega_{\mathrm{DE}}$, 
$\Omega_{\mathrm{m}} \gg \Omega_{\mathrm{r}}$) follows, 
and eventually dark energy becomes 
dominant over matter for $z<z_{\mathrm{DE}}$. 
Explicitly, we have $z_{\mathrm{DE}} = 0.36$, $0.40$ and $0.49$ for 
$u=1$, $0.8$ and $0.5$, respectively. 
As a consequence, 
after the radiation and matter-dominated stages, 
the current accelerated expansion of the universe 
can be realized in 
the combined $f(T)$ theory 
in Eq.~(\ref{eq:4.3}), 
similar to that in the exponential $f(T)$ theory.

Finally, we remark that since the gravitational field equation in $f(R)$ 
gravity 
has the fourth-order nature, 
it is possible to reproduce an arbitrary background expansion history, 
provided that we do not consider the stability 
issue~\cite{Song:2006ej,Dunsby:2010wg}). 
On the other hand, 
in the modified Gauss-Bonnet or $f(G)$ gravities, 
the $\Lambda$CDM background expansion history cannot be achieved 
all the way down to the far future~\cite{Li:2007jm}. 
In $f(T)$ gravity, there exists a similar restriction. 
Physically, in such modified gravity theories 
the modified background expansion rate $H$ is in general 
a function of $H$ itself and $\dot{H}$, 
which means that arbitrary expansion histories are not guaranteed to 
be realizable.

\section{Observational constraints on the combined $f(T)$ theory}

In Sec.\ IV,
we have constructed 
the combined $f(T)$ theory with both logarithmic and exponential terms 
to realize 
the crossing of the phantom divide.
Since 
it is difficult to obtain the information on the form 
of $f(T)$ from more fundamental theories,
it is important to examine 
whether this theory is compatible with the recent observational data.
In this section 
we demonstrate how well the combined $f(T)$ theory 
can fit the observational data. 
We examine the constraints on the 
model parameter $u$ and the current  
fractional non-relativistic matter density 
$\Omega_{\mathrm{m}}^{(0)}$ 
by taking the $\chi^{2}$ method for the recent observational data. 
We use 
type Ia supernovae (SNe Ia) data from 
the Supernova Cosmology Project (SCP) Union2 
compilation~\cite{Amanullah:2010vv}, 
the baryon acoustic oscillations (BAO) data from 
the Two-Degree Field Galaxy Redshift Survey (2dFGRS)
and the Sloan Digital Sky Survey Data Release 7 
(SDSS DR7)~\cite{Percival:2009xn}, 
and 
the cosmic microwave background (CMB) radiation data 
from Seven-Year Wilkinson Microwave Anisotropy Probe (WMAP)
observations~\cite{Komatsu:2010fb}. 
The detailed analysis method~\cite{Yang:2010xq} for the observational data of 
SNe Ia, BAO and CMB is summarized in Appendix A. 

\begin{center}
\begin{figure}[tbp]
\resizebox{!}{6.5cm}{
   \includegraphics{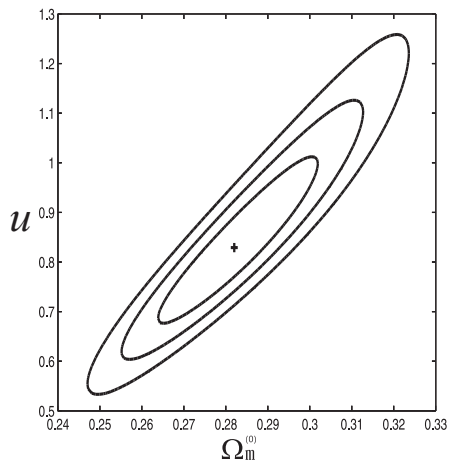}
                  }
\caption{
Contours of 
68.27\% ($1\sigma$), 95.45\% ($2\sigma$) and 99.73\% ($3\sigma$) 
confidence levels in the $(\Omega_{\mathrm{m}}^{(0)}, u)$ plane 
from SNe Ia, BAO and CMB data 
for the combined $f(T)$ theory, 
where the plus sign depicts the best-fit point. 
}
\label{fig-13}
\end{figure}
\end{center}

\begin{table*}[tbp]
\caption{
The best-fit values of $u$, 
$\Omega_{\mathrm{m}}^{(0)}$, 
$h$ and $\chi_{\mathrm{min}}^{2}$ 
for the combined $f(T)$ 
and $\Lambda$CDM models. 
}
\begin{center}
\begin{tabular}
{ccccc}
\hline
\hline
Model
& $u$
& $\Omega_{\mathrm{m}}^{(0)}$
& $h$
& $\chi_{\mathrm{min}}^{2}$
\\[0mm]
\hline
$f(T)$ 
& $0.829$
& $0.282$
& $0.691$
& $544.56$ 
\\[1mm]
$\Lambda$CDM 
& 
& $0.275$
& $0.707$
& $545.23$
\\[1mm]
\hline
\hline
\end{tabular}
\end{center}
\label{table-1}
\end{table*}

We plot the contours of 
the 68.27\% ($1\sigma$), 95.45\% ($2\sigma$) and 99.73\% ($3\sigma$)  
confidence levels (CL) in the $(\Omega_{\mathrm{m}}^{(0)}, u)$ plane 
from the SNe Ia, BAO and CMB data 
for the combined $f(T)$ theory
in Fig.~\ref{fig-13}. 
In this figure, 
we have fitted three parameters of $u$, $\Omega_{\mathrm{m}}^{(0)}$ 
and $h \equiv H_{0}/100/[\mathrm{km} \, \mathrm{sec}^{-1} \, 
\mathrm{Mpc}^{-1}]$~\cite{Kolb and Turner}. 
The model parameter $u$ is constrained to 
$0.6<u<1.13$ (95.45\% CL) 
whereas 
$0.255< \Omega_{\mathrm{m}}^{(0)} <0.312$ (95.45\% CL), 
which is consistent with the current observations. 
The best-fit value, i.e., 
the minimum $\chi^{2}$, in the parameter space is 
$\chi_{\mathrm{min}}^{2} = 544.56$ with $u = 0.829$, 
$\Omega_{\mathrm{m}}^{0} = 0.282$ and $h= 0.691$. 
In Table~\ref{table-1}, we present 
the best-fit values of $u$, 
$\Omega_{\mathrm{m}}^{(0)}$ and 
$h$, and $\chi_{\mathrm{min}}^{2}$ 
for the combined $f(T)$ theory along with 
those for the $\Lambda$CDM model. 
{}From Table~\ref{table-1}, we see that 
$\chi_{\mathrm{min}}^{2}$ 
of the combined $f(T)$ theory is slightly smaller than that of 
the $\Lambda$CDM model. 
This implies that the combined $f(T)$ theory can fit the observational data 
well. 

We remark that for around the best-fit value of the model parameter 
$u \approx 0.8$, 
$w_{\mathrm{DE}}$ can cross the phantom divide line of $w_{\mathrm{DE}}=-1$ 
from the non-phantom phase ($w_{\mathrm{DE}} > -1$) to 
the phantom one ($w_{\mathrm{DE}} < -1$) 
in the combined $f(T)$ theory as shown in Fig.~\ref{fig-7}, 
and therefore 
the observational data may favor the crossing of the phantom divide. 
This consequence agrees with the one in Ref.~\cite{Wu:2010av}.

\section{conclusions}

We have investigated the cosmological evolution in 
the exponential $f(T)$ theory, which is one of the simplest 
$f(T)$ theories as it contains only one model parameter $p$ along with 
the current fractional density of non-relativistic matter 
$\Omega_{\mathrm{m}}^{(0)}$. 
We have explicitly shown that 
the phase of the universe 
depends on the sign of the parameter $p$, i.e., 
for $p<0(>0)$ the universe is always in the non-phantom (phantom) phase 
without the crossing of the phantom divide. 
We have presented another simplest $f(T)$ model, 
which is the logarithmic type. 
Similar to the exponential model, the logarithmic one has only one free 
parameter $q$, and it does not allow the crossing of the phantom divide. 
To realize the crossing of the phantom divide, 
we have constructed a $f(T)$ theory by combining the 
logarithmic and exponential terms. 
In particular, we have shown that the crossing in the combined $f(T)$ theory 
is from $w_{\mathrm{DE}} > -1$ to $w_{\mathrm{DE}} < -1$, 
which is opposite to the typical manner in $f(R)$ gravity models. 
Furthermore, we have also illustrated that this combined theory is consistent 
with the recent observational data of SNe Ia, BAO and CMB. 

\section*{Acknowledgments}

One of the authors (CQG) would like to thank
the Institute of Theoretical Physics, Beijing for hospitality and a wonderful 
program on ``Dark Energy and Dark Matter'' at Weihai, 
and Professors Rong-Gen Cai, Yue-Liang Wu and Hongwei Yu for 
useful discussions. 
The work is supported in part by 
the National Science Council of R.O.C. under
Grant \#: 
NSC-98-2112-M-007-008-MY3
and 
National Tsing Hua University under the Boost Program \#: 
99N2539E1.

\appendix
\section{Analysis method for the observational data}

In this appendix, we explain the analysis method~\cite{Yang:2010xq} 
for the observational data of 
type Ia supernovae (SNe Ia), 
baryon acoustic oscillations (BAO) 
and cosmic microwave background (CMB) radiation 
(for more detailed explanations on the data analysis, 
see, e.g.,~\cite{Li:2009jx}).

\subsection{Type Ia Supernovae (SNe Ia)}

The information on the luminosity distance $D_{L}$ as a function
of the redshift $z$ is given by SNe Ia observations. 
The theoretical distance modulus $\mu_{\mathrm{th}}$ is defined by 
\begin{equation}
\mu_{\mathrm{th}}(z_{i})\equiv5\log_{10}D_{L}(z_{i})+\mu_{0}\,, 
\label{eq:A.1} 
\end{equation}
where $\mu_{0}\equiv42.38-5\log_{10}h$ with 
$h \equiv H_{0}/100/[\mathrm{km} \, \mathrm{sec}^{-1} \, 
\mathrm{Mpc}^{-1}]$~\cite{Kolb and Turner}. 
The Hubble-free luminosity
distance for the flat universe is described as 
\begin{equation}
D_{L}(z)=\left(1+z\right)\int_{0}^{z}\frac{dz'}{E(z')}\,,
\label{eq:A.2} 
\end{equation} 
where $E(z) \equiv H(z)/H_{0}$ 
with 
\begin{equation} 
H(z)=H_{0}
\sqrt{\Omega_{\mathrm{m}}^{(0)}\left(1+z\right)^{3}
+\Omega_{\mathrm{r}}^{(0)}\left(1+z\right)^{4}
+\Omega_{\mathrm{DE}}^{(0)}\left(1+z\right)^{3\left(1+w_{\mathrm{DE}}\right)}
}\,. 
\label{eq:A.3} 
\end{equation} 
Here, 
$\Omega_{\mathrm{r}}^{(0)}=\Omega_{\gamma}^{(0)}
\left(1+0.2271N_{\mathrm{eff}}\right)$, 
where $\Omega_{\gamma}^{(0)}$ is the present fractional photon energy density 
and $N_{\mathrm{eff}}=3.04$ is the effective number of neutrino 
species~\cite{Komatsu:2010fb}. 
We note that $H(z)$ is evaluated by using numerical solutions of 
Eq.~(\ref{eq:3.3}). 

The $\chi^{2}$ of the SNe Ia data is given by 
\begin{equation}
\chi_{\mathrm{SN}}^{2}=\sum_{i}\frac{\left[\mu_{\mathrm{obs}}(z_{i})-
\mu_{\mathrm{th}}(z_{i})\right]^{2}}{\sigma_{i}^{2}}\,,
\label{eq:A.4} 
\end{equation} 
where $\mu_{\mathrm{obs}}$ is the observed value of the distance modulus. 
In what follows, subscriptions ``th" and ``obs" denote 
the theoretically predicted and observed values, respectively. 
$\chi_{\mathrm{SN}}^{2}$ should be minimized 
with respect to $\mu_{0}$, 
which relates to the absolute magnitude, 
because the absolute magnitude of SNe Ia is not known. 
$\chi_{\mathrm{SN}}^{2}$ in Eq.~(\ref{eq:A.4}) is expanded to be~\cite{CS-SN} 
\begin{equation}
\chi_{SN}^{2}=A-2\mu_{0}B+\mu_{0}^{2}C\,,
\label{eq:A.5} 
\end{equation} 
with 
%
\begin{equation}
A 
= 
\sum_{i}\frac{\left[\mu_{\mathrm{obs}}(z_{i})-\mu_{\mathrm{th}}(z_{i};\mu_{0}=0)\right]^{2}}{\sigma_{i}^{2}}\,, 
\quad 
B 
=
\sum_{i}\frac{\mu_{\mathrm{obs}}(z_{i})-\mu_{\mathrm{th}}(z_{i};\mu_{0}=0)}
{\sigma_{i}^{2}}\,, 
\quad 
C=\sum_{i}\frac{1}{\sigma_{i}^{2}}\,.
\label{eq:A.6} 
\end{equation} 
The minimum of $\chi_{\mathrm{SN}}^{2}$ with respect to $\mu_{0}$ is expressed 
as 
\begin{equation}
\tilde{\chi}_{\mathrm{SN}}^{2}=A-\frac{B^{2}}{C}\,. 
\label{eq:A.7} 
\end{equation} 
In our analysis, we take Eq.~(\ref{eq:A.7}) for 
the $\chi^{2}$ minimization 
and use 
the Supernova Cosmology Project (SCP) Union2 compilation, which contains 557 
supernovae~\cite{Amanullah:2010vv}, ranging from $z=0.015$ to $z=1.4$.

\subsection{Baryon Acoustic Oscillations (BAO)}

The distance ratio of $d_{z}\equiv r_{s}(z_{\mathrm{d}})/D_{V}(z)$
is measured by the observation of BAO. 
Here, $D_{V}$ is the volume-averaged distance, $r_{s}$ is the comoving
sound horizon and $z_{\mathrm{d}}$ is the redshift 
at the drag epoch~\cite{Percival:2009xn}. 
The volume-averaged distance $D_{V}(z)$ is defined as~\cite{BAO}
\begin{equation}
D_{V}(z)\equiv\left[\left(1+z\right)^{2}
D_{A}^{2}(z)\frac{z}{H(z)}\right]^{1/3}\,,
\label{eq:A.8} 
\end{equation} 
where $D_{A}(z)$ is the proper angular diameter distance for the flat 
universe, defined by 
\begin{equation}
D_{A}(z) \equiv 
\frac{1}{1+z}\int_{0}^{z}\frac{dz'}{H(z')}\,. 
\label{eq:A.9} 
\end{equation} 
The comoving sound horizon $r_{s}(z)$ is given by
\begin{equation}
r_{s}(z)=\frac{1}{\sqrt{3}}\int_{0}^{1/(1+z)}\frac{da}{a^{2}
H(z^{\prime}=1/a-1) \sqrt{1+\left(3\Omega_{b}^{(0)}/4\Omega_{\gamma}^{(0)}
\right)a}}\,,
\label{eq:A.10} 
\end{equation} 
where $\Omega_{b}^{(0)} = 2.2765 \times10^{-2} h^{-2}$ and 
$\Omega_{\gamma}^{(0)} = 2.469\times10^{-5}h^{-2}$ are the current 
values of baryon and photon density parameters, 
respectively~\cite{Komatsu:2010fb}. 
The fitting formula for $z_{\mathrm{d}}$ is given 
by~\cite{Eisenstein:1997ik} 
\begin{equation}
z_{\mathrm{d}}=\frac{1291(\Omega_{\mathrm{m}}^{(0)}h^{2})^{0.251}}
{1+0.659(\Omega_{\mathrm{m}}^{(0)}h^{2})^{0.828}}\left[1+b_{1}
\left(\Omega_{b}^{(0)}h^{2}\right)^{b2}\right]\,,
\label{eq:A.11} 
\end{equation} 
with 
%
\begin{equation}
b_{1}=0.313(\Omega_{\mathrm{m}}^{0}h^{2})^{-0.419}\left[
1+0.607\left(\Omega_{\mathrm{m}}^{0}h^{2}\right)^{0.674}\right]\,, \quad 
b_{2}=0.238\left(\Omega_{\mathrm{m}}^{0}h^{2}\right)^{0.223}\,.
\label{eq:A.12} 
\end{equation} 
The typical value of $z_{\mathrm{d}}$ is about $1021$ 
for $\Omega_{\mathrm{m}}^{(0)}=0.276$ and $h=0.705$. 

According to the BAO data from 
the Two-Degree Field Galaxy Redshift Survey (2dFGRS)
and the Sloan Digital Sky Survey Data Release 7 
(SDSS DR7)~\cite{Percival:2009xn}, 
the distance ratio $d_{z}$ at two redshifts $z=0.2$ and 
$z=0.35$ 
is measured to be $d_{z=0.2}^{\mathrm{obs}}=0.1905\pm0.0061$ and 
$d_{z=0.35}^{\mathrm{obs}}=0.1097\pm0.0036$
with the inverse covariance matrix: 
%
\begin{equation}
C_{\mathrm{BAO}}^{-1}=\left(
\begin{array}{cc}
30124 & -17227\\
-17227 & 86977
\end{array}\right)\,.
\label{eq:A.13} 
\end{equation} 
%
The $\chi^{2}$ for the BAO data is described as 
%
\begin{equation}
\chi_{\mathrm{BAO}}^{2}=
\left(x_{i,\mathrm{BAO}}^{\mathrm{th}}-x_{i,\mathrm{BAO}}^{\mathrm{obs}}\right)
\left(C_{\mathrm{BAO}}^{-1}\right)_{ij}
\left(x_{j,\mathrm{BAO}}^{\mathrm{th}}-x_{j,\mathrm{BAO}}^{\mathrm{obs}}
\right)\,,
\label{eq:A.14} 
\end{equation} 
%
where $x_{i,\mathrm{BAO}} \equiv \left(d_{0.2},d_{0.35}\right)$.

\subsection{Cosmic Microwave Background (CMB) radiation}

The observational data of CMB are sensitive to the distance 
to the decoupling epoch $z_{*}$~\cite{Komatsu:2008hk}. 
Hence, by using these data we obtain 
constraints on the model in the high redshift regime ($z\sim1000$). 

The acoustic scale $l_{A}$ and 
the shift parameter 
$\mathcal{R}$~\cite{Bond:1997wr} 
are defined by 
\begin{eqnarray}
l_{A}(z_{*})
\Eqn{\equiv}
\left(1+z_{*}\right)\frac{\pi D_{A}(z_{*})}{r_{s}(z_{*})}\,,
\label{eq:A.15} \\ 
\mathcal{R}(z_{*})
\Eqn{\equiv}
\sqrt{\Omega_{\mathrm{m}}^{(0)}}H_{0}
\left(1+z_{*}\right)D_{A}(z_{*})\,, 
\label{eq:A.16} 
\end{eqnarray}
where $z_{*}$ is the redshift of the decoupling epoch, 
given by~\cite{Hu:1995en}
\begin{equation}
z_{*}=1048\left[1+0.00124\left(\Omega_{b}^{(0)}h^{2}\right)^{-0.738}\right]\left[1+g_{1}\left(\Omega_{\mathrm{m}}^{(0)}h^{2}\right)^{g2}\right]\,,
\label{eq:A.17} 
\end{equation} 
with
\begin{equation}
g_{1}=\frac{0.0783\left(\Omega_{b}^{(0)}h^{2}\right)^{-0.238}}{1+39.5\left(\Omega_{b}^{(0)}h^{2}\right)^{0.763}},\quad g_{2}=\frac{0.560}{1+21.1\left(\Omega_{b}^{(0)}h^{2}\right)^{1.81}}\,.
\label{eq:A.18} 
\end{equation} 
We use the 
data from Seven-Year Wilkinson Microwave Anisotropy Probe (WMAP)
observations~\cite{Komatsu:2010fb} on CMB. 

The $\chi^{2}$ of the CMB data is 
\begin{equation}
\chi_{\mathrm{CMB}}^{2}=\left(x_{i,\mathrm{CMB}}^{\mathrm{th}}-x_{i,\mathrm{CMB}}^{\mathrm{obs}}\right)
\left(C_{\mathrm{CMB}}^{-1}\right)_{ij}
\left(x_{j,\mathrm{CMB}}^{\mathrm{th}}-x_{j,\mathrm{CMB}}^{\mathrm{obs}}
\right)\,,
\label{eq:A.19} 
\end{equation} 
where $x_{i,\mathrm{CMB}}\equiv\left(l_{A}(z_{*}), 
\mathcal{R}(z_{*}), z_{*}\right)$
and $C_{\mathrm{CMB}}^{-1}$ is the inverse covariance matrix. 
The data from WMAP7 observations~\cite{Komatsu:2010fb} 
lead to $l_{A}(z_{*})=302.09$,
$\mathcal{R}(z_{*})=1.725$ and $z_{*}=1091.3$ with the inverse covariance 
matrix: 
\begin{equation}
C_{\mathrm{CMB}}^{-1}=\left(\begin{array}{ccc}
2.305 & 29.698 & -1.333\\
29.698 & 6825.27 & -113.180\\
-1.333 & -113.180 & 3.414\end{array}\right)\,.
\label{eq:A.20} 
\end{equation} 

Consequently, 
the $\chi^{2}$ of all the observational data is given by
\begin{equation}
\chi^{2}=\tilde{\chi}_{\mathrm{SN}}^{2}+\chi_{\mathrm{BAO}}^{2}+\chi_{\mathrm{CMB}}^{2}\,.
\label{eq:A.21} 
\end{equation} 
We note that 
in our fitting procedure, 
we take the simple $\chi^{2}$ method rather than 
the Markov chain Monte Carlo (MCMC) approach such as 
CosmoMC~\cite{Lewis:2002ah}.


\end{document}